\documentclass[11pt,3p,authoryear,preprint]{elsarticle}
\usepackage[T1]{fontenc}
\usepackage{lineno,hyperref}
\usepackage{eurosym}
\usepackage{booktabs}
\usepackage{multirow}
\usepackage{mathrsfs}
\usepackage{graphicx}         
\usepackage{multicol}       
\usepackage{yfonts}
\usepackage{xcolor}
\usepackage{nicefrac}
\usepackage[linesnumbered,ruled,vlined]{algorithm2e}
\usepackage{amsmath,dsfont}
\usepackage{float}
\usepackage{natbib}
\usepackage[section]{placeins}
\usepackage{amssymb}
\usepackage{dsfont}
\usepackage{mathabx}
\usepackage[toc,page]{appendix}
\newtheorem{defi}{Definition}

\newtheorem{remark}{Remark}
\newtheorem{theorem}{Theorem}
\newtheorem{bsp}{Example}
\newtheorem{cor}{Corollary}
\newtheorem{prop}{Proposition}
\newtheorem{as}{Assumption}

\usepackage{booktabs}
\usepackage{siunitx}
\usepackage{array}
\usepackage{caption}
\usepackage{listings}

\usepackage{textcomp}
\usepackage{xcolor}

\definecolor{codegreen}{rgb}{0,0.6,0}
\definecolor{codegray}{rgb}{0.5,0.5,0.5}
\definecolor{codepurple}{rgb}{0.58,0,0.82}
\definecolor{backcolour}{rgb}{0.95,0.95,0.92}

\lstdefinestyle{mystyle}{ 
    basicstyle=\ttfamily\footnotesize,
    columns=fullflexible,
    backgroundcolor=\color{backcolour},   
    commentstyle=\color{codegreen},
    keywordstyle=\color{magenta},
    numberstyle=\tiny\color{codegray},
    stringstyle=\color{codepurple},
    basicstyle=\ttfamily\footnotesize,
    breakatwhitespace=false,         
    breaklines=true,    captionpos=b,                    
    keepspaces=true, numbers=left,                    
    numbersep=5pt, showspaces=false,                
    showstringspaces=false, showtabs=false, tabsize=2}
\SetCommentSty{mycommfont}

\SetKwInput{KwInput}{Input}                
\SetKwInput{KwOutput}{Output}              
\journal{}
\begin{document}
\allowdisplaybreaks
\begin{frontmatter}

\title{Empirical Decision Theory}

\author[rvt]{C.~Jansen\corref{cor1}\fnref{equal}}
\ead{c.jansen@lancaster.ac.uk}
\author[rvt1]{G.~Schollmeyer\fnref{equal}}
\author[rvt1]{T.~Augustin}
\author[rvt1,rvt2]{J.~Rodemann}
\cortext[cor1]{Corresponding author}
\fntext[equal]{These authors contributed equally to this work.}
\address[rvt]{School of Computing \& Communications, Lancaster University Leipzig, Germany}
\address[rvt1]{Department of Statistics, LMU Munich, Germany}
\address[rvt2]{CISPA Helmholtz Center for Information Security, Saarbrücken, Germany}

\begin{abstract}
Analyzing decision problems under uncertainty commonly relies on idealizing assumptions about the describability of the world, with the most prominent examples being the \textit{closed world} and the \textit{small world assumption}. Most assumptions are operationalized by introducing \textit{states of the world}, conditional on which the decision situation can be analyzed without any remaining uncertainty. Conversely, most classical decision-theoretic approaches are not applicable if the states of the world are inaccessible. We propose a decision model that retains the appeal and simplicity of the original theory, but completely overcomes the need to specify the states of the world explicitly. The main idea of our approach is to address decision problems in a radically empirical way: instead of specifying states and consequences prior to the decision analysis, we only assume a protocol of observed act--consequence pairs as model primitives. We show how optimality in such \textit{empirical decision problems} can be addressed by using protocol-based \textit{empirical choice functions }and discuss three approaches for deriving inferential guarantees: (I) consistent statistical estimation of choice sets, (II) consistent statistical testing of choice functions with robustness guarantees, and (III) direct inference for empirical choice functions using credal sets. We illustrate our theory with a proof-of-concept application comparing different prompting strategies in generative AI models.
\end{abstract}


\end{frontmatter}

\section{Introduction}
\noindent This article deals with \textit{\textbf{empirical} decision problems (or shorthand: EDPs)}. By this term, we refer to decision problems under (generalized) uncertainty whose possible states of the world are \textit{not a priori available to the decision maker}, but can only be made accessible indirectly via appropriate sampling strategies (e.g., by collecting the consequences different decisions lead to in past situations that can be judged to be similar enough to the present one). In other words, rather than assuming to know all the states of the world and (equally restrictive) all the consequences that different decisions under a given state would theoretically lead
to, our model primitive is a so-called \textit{protocol}, i.e., an (observed) set of act-consequence pairs, each specifying what consequence a certain decision led to when performed in a past situation under equal (or very similar) conditions. We will repeatedly refer to the following example to provide our theory with an easy-to-capture intuition.
\begin{bsp} \label{mot_ex}
For motivational purposes, a company has established the following innovation to its time recording system: Each employee can choose to clock out with one of five different cards, each having a different color, say red, blue, green, yellow, and black. After clocking out, some amount of money is loaded onto the employee's canteen payment card. The clock-out machine is set to an unknown status every morning according to an unknown probability distribution (or some other unknown mechanism). Suppose an employee has experimented with the different cards to clock out over a whole month and documented the amount received each time. Can a good choice strategy for the next month be derived from this information? 
\end{bsp}

\subsection{Motivation and Background}
\noindent The following questions come to mind: Why is a modification of the classical framework of decision theory under uncertainty\footnote{We will be more precise about what we mean by the \textit{classical} framework in Sections~\ref{classical1} and~\ref{classical}.} interesting or even necessary? Hasn't the classical approach achieved considerable success and produced a mathematically and philosophically beautiful theory?~Of course, this is the case, and this paper does not seek to cast any doubt on that whatsoever. However, we believe that there are (many) relevant decision-making situations that simply cannot be analyzed within the traditional framework. Specifically, we will argue that by transitioning from classical decision problems to empirical ones, we are able to overcome two major challenges of classical decision theory: The \textit{small world problem} referring to the (potentially) arbitrary set granularity of the state space under consideration (and the dependence of an optimal decision on this setting), and the \textit{closed world problem}, referring directly to the practical lack of ability of specifying an suitable state space. Even more, we will argue that specifying an appropriate state space is very hard in many (if not most) \textit{practical} decision situations, emphasizing that the problems just mentioned are far from being only theoretical and, thereby, making the possibility to overcome them even more attractive. Or, in the words of the decision theory pioneer Edi Karni, we can attest that \textit{``[...]~careful examination of the concept of state of the world reveals that the depiction of the relevant state-space is often unintuitive and too complex to be compatible with the decision makers' perception of choice problems''} \citep[p.~325]{KARNI2006325}.
\subsection{Related Work}
\noindent Given the generality and importance of the problem of inaccessible state spaces, it hardly surprises that there can be found many different approaches for tackling it in the literature. In our eyes, the following among these approaches seem particularly prominent and relevant for our discussion: The first is Gilboa and Schmeidler's \textit{case-based} approach to decision making (see  \cite{gilboa1995case,Gilboa_Schmeidler_2001,billot2008axiomatization}), where decisions -- rather than comparing their expected utilities -- are evaluated by a weighted sum of the utilities they have produced in similar problems in the agent's memory. Similarity here is measured by a function assigning a real-valued score to any pair of problems. While the aforementioned works focus in particular on the axiomatic characterization of this alternative decision theory, subsequent works such as \cite{hu1998,b2017} have shown how similarity measures can be elicited and thus built a bridge to practical applications such as \cite{guerdjikova2004case,lovallo2012robust,grosskopf2015experiment}. 
\\[.2cm]
The second is Shafer's \textit{constructive} approach to decision theory (see \cite{s2016} or, somehow related, also \cite{Kaneko2019,sa2023}), where -- instead of states -- the agent specifies a list of goals and then judges the quality of a decision by its compatibility with those pre-defined goals. From a technical point of view, the mathematical machinery here is mainly based on the theory of belief functions \citep{s1976,denoeux2019decision}. To the best of our knowledge, Shafer's constructive decision theory has not yet been extensively incorporated into applied research. However, it shows interesting parallels to literature on human-AI alignment, where the question also arises as to how well a decision-making algorithm provided by an AI can fulfill the goals articulated by a human. Some recent works in this directions are \cite{pmlr-v235-yang24p,pos-25}.
\\[.2cm]
A third approach is Blume, Easley, and Halpern's \textit{constructive} approach to decision making (see \cite{BLUME2021105306}), where (essentially) decisions (or acts) are replaced by \textit{syntactic programs} (simple if-else rules) and states are replaced by propositions about the world. These model primitives are then used to \textit{derive} (or \textit{construct}) acts, states, and consequences corresponding to a classic Savageian setting. More precisely, the authors prove here an existence result stating that if the preferences of the decision-maker over these programs satisfy certain axioms, then the information from these preferences can be used to construct a classic decision problem.
\\[.2cm]
Finally, there is a number of technically deep papers that describe rational decision-making axiomatically without having to specify a state space or find ways to do it in a justified manner: In \cite{dekel2001representing}, the authors provide an axiomatic characterization of preference structures that enables the construction of an (essentially) unique \textit{subjective} state space. The paper by \cite{ghirardato2001coping} introduces a way to formally represent situations where the decision maker cannot list all possible states, and show how such situations under \textit{unforeseen contingencies} can be tackled by using Choquet expected utility rather than classical one. Moreover, while \cite{machina2003states} gives a critical reflection of the role of states of the world in decision theory and surveys and discusses different approaches avoiding the use of states, the papers \cite{KARNI2006325} and \cite{a2008} manage to axiomatically recover subjective expected utility theory and maximin expected utility theory, respectively, without ever referring to a state space at all. Finally, in \cite{KARNI2022105444} acts (and their respective outcomes) are interpreted as experiments that are used to provide evidence about the validity of different competing theories about which act would lead to which outcome. States are replaced by such theories in this model.
\\[.2cm]
\noindent Interestingly, what almost all of these works have in common is that they propose \textit{alternative models} to Savage's classic model, rather than providing solutions to the challenges \textit{within} this framework. More precisely, all these models address the problem of the non-specifiable state space by replacing it with a constructive, more easily specifiable component, thereby, potentially losing what made Savage's framework of comparing acts mapping states to consequences so appealing at first. Against this background, we should also explicitly stress that we, in particular, aim to maintain the prescriptive character of Savage's theory, informing us about how a rational agent ought to decide.  Thus, our notion of an {\em empirical} decision theory must not be confused with concepts of {\em descriptive} decision theory that, in the tradition of \cite{Simon:1955:QJE} and \cite{Tversky:Kahneman:1974:Science}, collect and systematize empirical evidence on how decisions with their heuristics and biases are concretely made under bounded rationality.\\[2mm]
For our robustness guarantees in Section~\ref{testing} and~\ref{direct_inference}, we will heavily rely on contamination models. They originate in robust statistics \citep{Huber:1965,Huber:1981,Huber:2009}.  Early treatments offered robust Bayesian frameworks \citep{DeRobertis:1981}, while more recent works comprise contamination models often under the term `linear-vacuous mixture' within the general theory of imprecise probabilities based on credal sets and coherent lower previsions (\cite{Walley:1991,augustin2014introduction,Troffaes:2014} for monographs and \cite{ISIPTA:2023,ISIPTA:2025} for recent special proceedings). In this view, contamination models formalize varying degrees of ambiguity 
by mixing a nominal distribution to an appropriate extent with alternative distributions. A central example is the Imprecise Dirichlet Model \citep{walley1996inferences} for inference with multinomial data, which admits a contamination-interpretation based on the observed relative frequencies \citep{Seidenfeld:Wassermann:on:Walley:1996}. For a comprehensive exposition of credal inference and updating, we refer the interested reader again to \citet{Walley:1991, augustin2014introduction,Troffaes:2014}. In the (robust) Bayesian literature, contamination models remain a canonical device for global sensitivity analysis and have been applied and extended in various ways \citep{Berger:1994,mangili2015new,mangili2016prior,benavoli2015prior,rodemann2021accounting,rodemann2022accounting,rodemann2024imprecise,meets,festschrift,capriocredal}.
%
%
\subsection{Contribution and Outline}
\noindent In contrast to the alternatives decision theoretic approaches just discussed, we want to take a path here that retains the appeal and simplicity of the original theory, but avoids the necessity to specify the states of the world explicitly. The main idea of our approach is to address decision problems in a radically empirical way: instead of specifying states and consequences a priori to the decision analysis, we assume our decision problem to consist only of \textit{observable} components of the form \textit{act $\rightarrow$ consequence}, i.e.,~of a protocol of act-consequence-pairs storing the information which act led to which consequence in previous trials of choosing the corresponding act (see Definition~\ref{def:sdp} of Section~\ref{sec:edp}). Based on these protocols, we go on defining so-called \textit{empirical choice functions} (see Definition~\ref{def:ecf} of Section~\ref{sec:edp}). Like classical choice functions, these assign to each potential input a set of decisions that appear to be appropriate choices under that input. The difference lies in the domain: While classical choice functions operate on a set system over the acts (i.e., \textit{require} knowledge of the state space), empirical choice functions refer solely to the available protocol. No explicit knowledge of the states is required. Note that, while we \textit{do} need to assume the existence of a (latent) state space and that every decision option can be identified with a latent act on this space, our assumption is nevertheless substantially weaker than that of the classical theory, which presumes that these can be made explicit. In plain terms, our assumption essentially is that the decision maker believes in a non-deterministic process generating the states of the world.
\\[.2cm]
\noindent
So, how should the decision sets select by an empirical choice function be interpreted? We provide two answers: First, we show under which conditions an empirical choice function can converge towards its population variant (i.e., the choice function whose empirical analog it models). Under these conditions, an empirical choice function can then be interpreted as a consistent statistical estimator for its population analog (see Theorem~\ref{theorem_consistent_approximator} of Section~\ref{estimation}). Second, we show how to construct an appropriate statistical test for checking whether some act is contained in the choice set of the population variant of an empirical choice function under investigation (see Theorem~\ref{theo2} of Section~\ref{testing}). From this angle, the choice sets of the underlying empirical choice function can then be interpreted as being statistically significant candidates for the true choice sets under some pre-specified error level. Further, we show how the proposed statistical test can be modified to remain valid even under slight deviations from its main underlying assumption, namely the \textit{i.i.d.} structure of the observed consequences for each action description belonging to the respective protocol (see Theorem~\ref{theo3} of Section~\ref{testing}). In Section~\ref{direct_inference}, we propose an alternative way to obtain inferential guarantees for solutions of empirical decision problems that we term \textit{direct credal inference}. We connect this to the results obtained in the classical inferential setting (see Corollary~\ref{cor:direct} of Section~\ref{direct_inference}). In Section~\ref{application}, we illustrate our theory with a proof-of-concept application comparing different prompting strategies in generative AI models. To make a discussion of EDPs and corresponding approaches for their solution possible, we begin by outlining the classical decision-theoretic setting on which we base this paper (Section~\ref{known_states}) and by providing some details of its limitation (Section~\ref{cwaswa}).
%
%
\section{The Classical Setting: Decision Problems with Known State Spaces} \label{known_states}
\noindent We now give a short overview of the concepts from decision making with \textit{known} state spaces which are relevant for building our theory in the later parts of this article. This section is in parts influenced by the presentation in \cite{jansen2018some,jansen2025contributionsdecisiontheoreticfoundations}. 
\subsection{Decision Making under Uncertainty}\label{classical1}
\noindent The basic formalism of \textit{decision making under uncertainty} is as simple as it is expressive: the \textit{agent} is asked to choose between different available \textit{acts} $X$ from a known set of acts $\mathcal{G}$. However, the \textit{consequence} of choosing an act $X$ is not (necessarily) deterministic, but rather depends on which \textit{state of the world} from a \textbf{known} set $S$ of such states turns out to be the true one. Formally, each act is a mapping $X:S \rightarrow C$, where $C$ is the set of all possible consequences. The \textit{decision problem} $\mathcal{G}$ is then some fixed subset of the set of all possible acts, i.e., the set $C^S=\{X:S \to C\}$. 
\\[.2cm]
\noindent The agent's goal is to select an optimal (set of) act(s) or, more general, a set of acts for which there is no evidence that one is in some sense superior to the others. This selection process is commonly formalized by a \textbf{\textit{choice function}} $ch:2^{\mathcal{G}} \to 2^{\mathcal{G}}$ satisfying $ch(\mathcal{D}) \subseteq \mathcal{D} $ for all $\mathcal{D} \in 2^{\mathcal{G}}$.\footnote{Note that we also allow for empty choice sets, even if these are sometimes ruled out (for technical reasons) when discussing choice functions. Nothing hinges on this for the results of the article at hand.} The sets $ch(\mathcal{D})$ are called \textbf{\textit{choice sets}} and have a slightly different interpretation depending on the quality of the information used to construct the choice function: The \textit{strong view} interprets $ch(\mathcal{D})$ as the set of \textit{optimal} acts from $\mathcal{D}$. The \textit{weak view}, on the other hand, interprets $ch(\mathcal{D})$ as the set of acts from $\mathcal{D}$ that \textit{cannot be rejected} based on the information. Usually in this paper we will rather be concerned with only weakly interpretable choice functions.
\subsection{Two Extreme (Yet Classical) Choice Functions}\label{classical}
\noindent Usually choice functions are constructed relying on two building blocks: the information about the \textit{uncertainty mechanism} generating the states of the world in $S$, and the information about the agent's \textit{preferences} among the consequences in $C$. We will refer to these information bases as $\mathcal{I}_1$ and $\mathcal{I}_2$, respectively. For further reference, we now briefly recall two popular choice functions:
\\[.3cm]
\noindent
\textbf{Expected Utility (EU):} If the information in $\mathcal{I}_1$ allows to describe the uncertainty about the states by a (subjective) probability $\pi$ on $S$, and the information in $\mathcal{I}_2$ allows for expressing the agent's preferences by a cardinal (measurable) \textit{utility function} $u:C \to [0,1]$, then we can set
\begin{equation} \label{EU}
ch_{u,\pi}(\mathcal{D})=\Bigl\{Y \in \mathcal{D}: \mathbb{E}_\pi(u \circ Y) \geq \mathbb{E}_\pi(u \circ X) \text{ for all }X \in \mathcal{D}\Bigr\},
\end{equation}
for all $\mathcal{D} \subseteq \mathcal{G}$ and choose that acts from $\mathcal{D}$ that \textit{maximize expected utility}. The choice function $ch_{u,\pi}$ then obviously allows for a strong interpretation: all acts in $ch_{u,\pi}(\mathcal{D})$ produce coinciding expected utilities, which are strictly greater than all expected utilities of acts in $\mathcal{D} \setminus ch_{u,\pi}(\mathcal{D})$. Consequently, the agent is \textit{indifferent} among the equally optimal acts contained in $ch_{u,\pi}(\mathcal{D})$.
\\[.3cm]
\noindent
\textbf{First-Order Stochastic Dominance (FSD):} If the information in $\mathcal{I}_1$ allows to describe the uncertainty about the states by a (subjective) probability $\pi$ on $S$, and the information in $\mathcal{I}_2$ consists only in a preorder $\succsim$ on $C$, we set
\begin{equation} \label{FSD}
ch_{\succsim,\pi}(\mathcal{D})=\Biggl\{Y: \nexists X \begin{array}{lr} \forall u \in \mathcal{U}_{\succsim}:\mathbb{E}_\pi(u \circ X-u \circ Y)\geq 0\\\exists u \in \mathcal{U}_{\succsim}: \mathbb{E}_\pi(u \circ X-u \circ Y)>0
\end{array}\Biggr\}  
\end{equation}
for all $\mathcal{D} \subseteq \mathcal{G}$, where $\mathcal{U}_{\succsim}$ is the set of all $\succsim$-isotone (and measurable) functions\footnote{It is easy to see that instead of $\mathcal{U}_{\succsim}$ one can also equivalently use the set $\mathcal{I}_\succsim$ of all measurable and isotone indicator functions $u:C \to \{0,1\}$. The reason is simply that an isotone and bounded measurable function $u$ can be represented as $u:= \lim\nolimits_{k\to \infty} \sum\nolimits_{i=1}^{k²} 1/k \cdot \mathds{1}_{u \geq i/k}$ where all $\mathds{1}_{u\geq i/k }$ are isotone and measurable. Therefore, for the example of $C=\mathbb{R}$ (equipped with the usual order $\geq$), stochastic dominance can be equivalently characterized via the cumulative distribution function or the decumulative distribution function, like usually done for example in the econometrics literature like, e.g., \cite{Barrett,Abadie}.} $u:C \to [0,1]$. Thus, we choose all acts that are \textit{not excluded by every compatible EU-maximizer}. The choice function $ch_{\succsim,\pi}$ then obviously allows only for a weak interpretation: among all acts in $ch_{\succsim,\pi}(\mathcal{D})$ the agent is either \textit{indifferent} or considers them \textit{incomparable}. The acts in $\mathcal{D} \setminus ch_{\succsim,\pi}(\mathcal{D})$ are all strictly stochastically dominated by at least one act from $ch_{\succsim,\pi}(\mathcal{D})$. \\[.3cm]
But in what sense are the two choice functions just described extreme? This is very easy to see in the case of the expected utility rule: it requires very precise knowledge on both the cardinal structure of preferences (in the form of a cardinal utility function) and the process that generates the states of the world (in the form of a known and perfectly stochastic mechanism). From the classic axiomatic perspective of decision theory, such precise knowledge could be achieved if the decision maker orders the acts in $C^S$ in accordance with Savage's axioms \citep{foundstat}, which are often considered quite restrictive in real-world decision problems (e.g., \cite{HumphreyKruse2024}). However, even when using FSD choice functions, very consistent knowledge of the probabilities of occurrence of the states is still necessary. Again, from an axiomatic point of view, such knowledge could be generated if the decision maker can order all bounded real-valued functions (``gambles'') with domain $S$ according to De Finetti's axioms \citep{definetti1974theory}; a restriction that is considered so serious that it has developed into a field of research in its own right (see e.g.~the recent review by \cite{Gilboa:2025:Review} on generalized decision theory), in fact also founding the behavioral approach to imprecise probabilities \citep{Walley:1991}. Accordingly, we take the next paragraph to recall choice functions relaxing these extremes.
\subsection{Choice Functions Based on Expected Utility Between FSD and EU}
\noindent If the information sources $\mathcal{I}_1$ and $\mathcal{I}_2$ cannot be described by a precise probability measure $\pi$ and (or) a cardinal utility function $u$, then neither of the choice functions from Equations~\eqref{EU} or \eqref{FSD} are applicable, since in both cases necessary components for their respective evaluation are missing. However, if $\mathcal{I}_1$ is relaxed to a \textit{credal set} $\mathcal{M}$, i.e., a set of suitable probabilities measures on $S$,\footnote{The term as well as the concept \textit{credal set} originally stems from \cite{levi1974} and was popularized and theoretically established in the work following \cite{Walley:1991}. Note that credal sets are only one of various possible types of imprecise probabilistic models. Others rely on concepts like linear partial information \citep{kofler}, appropriately interpreted belief functions \citep{s1976}, singleton-based probability intervals \citep{weichselberger:poehlmann}, prevision \citep{Walley:1991}, and general interval probability \citep{Weichselberger:2001}. Presumably, most of the results obtained in this article also hold for these alternative models.} and if $\mathcal{I}_2$ is relaxed to a set of suitable utility functions $\mathcal{F}$, one (essentially) can still use a modification of Equation~\eqref{FSD} to induce a well-defined choice function by quantifying over all possible combinations of functions. Here, assuming a credal set $\mathcal{M}$ rather than one specific measure $\pi$ can be interpreted as collecting all probability measures that are compatible, i.e., non-contradictory, with the information on the generation of states provided by source $\mathcal{I}_1$. Similarly, going from one cardinal utility function $u$ to a set $\mathcal{F}$ of such, can be viewed as collecting all cardinal utilities that are compatible with whatever information is provided by source $\mathcal{I}_2$. This idea is formalized in the following definition:
\begin{defi} \label{EUbased}
    Let $\mathcal{M}$ be a credal set on the states $s \in S$ and let $\mathcal{F}$ be a set of (uniformly) bounded and measurable functions $f:C \to \mathbb{R}$. Define the \textbf{$(\mathcal{F},\mathcal{M})$-preorder} $\succsim_{(\mathcal{F},\mathcal{M})}$ among acts $X,Y$ by:

$$X \succsim_{(\mathcal{F},\mathcal{M})}Y~:\Leftrightarrow ~\forall \pi \in \mathcal{M}~\forall f \in \mathcal{F}:~\mathbb{E}_{\pi}(f \circ X) \geq \mathbb{E}_{\pi}(f \circ Y)$$
Extract the \textbf{$(\mathcal{F},\mathcal{M})$-choice function} $ch_{(\mathcal{F},\mathcal{M})}$ by setting, for every $\mathcal{D} \subseteq \mathcal{G}$,

$$ch_{(\mathcal{F},\mathcal{M})}(\mathcal{D}):= \bigl\{X\in \mathcal{D}: \nexists Y \in \mathcal{D} \text{ s.t. } Y \succ_{(\mathcal{F},\mathcal{M})}X\bigl\},$$
where $\succ_{(\mathcal{F},\mathcal{M})}$ denotes the strict part of $\succsim_{(\mathcal{F},\mathcal{M})}$. If $\mathcal{M}=\{\pi\}$ for some probability measure $\pi$ that is clear from the context, we abbreviate $\succsim_{(\mathcal{F},\mathcal{M})}$ by $\succsim_{\mathcal{F}}$ and $ch_{(\mathcal{F},\mathcal{M})}$ by $ch_{\mathcal{F}}$.
\end{defi}
The basic idea behind the choice function $ch_{(\mathcal{F},\mathcal{M})}$ is very similar to the basic idea behind the FSD-based choice function from Equation~\eqref{FSD}: The basic idea of maximizing expected utility is retained, but must be adapted to the non-uniqueness of the utility scale. However, in contrast to FSD, we here need to additionally adapt for the non-uniqueness, or \textit{ambiguity}, of the uncertainty representation, i.e., we select all acts that are not strictly EU-dominated by any other act for every compatible utility scale in $\mathcal{F}$ \textit{\textbf{and}} every compatible uncertainty representation in $\mathcal{M}.$ Note that the similarity to the choice functions seen so far can also be captured more formally: Given the probability measure $\pi$ is clear from the context, the relationships $ch_{\mathcal{U}_{\succsim}}= ch_{\succsim , \pi}$ and $ch_{\{u\}}= ch_{u, \pi}$ apply, i.e., the extreme choice functions based on EU and FSD from Equations~\eqref{EU} and \eqref{FSD} can be represented as special cases of the choice function $ch_{\mathcal{F}}$. Furthermore, the choice function based on \textit{generalized stochastic dominance (GSD)}, which has received some attention recently \citep{jsa2018,jbas2022,mdai,uaiall,jnsa2023,jansen2024statistical}, can also be represented as a special case of the choice function $ch_{\mathcal{F}}$: Still under the assumption that the probability measure $\pi$ is clear from the context, we just have to choose the function class $\mathcal{F}$ from Definition \ref{EUbased} as the set of representations of a preference system (see, e.g., \citet[Definition 1]{jsa2018} for the details).

%
%
%
%
\section{Small Worlds, Closed Worlds and the Definition of States}\label{cwaswa}
\noindent As already briefly foreshadowed in the introduction and as is also directly evident by inspecting the basic  decision-theoretic concepts recalled in the previous section, the analysis (and even the mere definition) of decision problems under uncertainty makes use of (and therefore heavily depends on) idealizing assumptions about the \textit{describability} of the world, i.e.~about the availability and the structure of the states of the world $S$ as well as the consequences $C$. A classic decision problem cannot be formally described or solved unless the quality of the decisions can be clearly analyzed conditional  on some fixed and well-defined set of hypothetical worlds. Even more, it must also be possible to clearly state which decision would lead to which consequence in each of these hypothetical worlds. This implies in particular that a decision needs to be completely identifiable with the random variable describing it on the set of states of the world. Conversely, of course, this means that classical decision-theoretical approaches run into severe trouble if the set of states of nature in the specific problem under consideration is inaccessible (or not appropriately approximizable) and, therefore, these implicit assumptions are violated.\\[.2cm] 
\noindent We want to distinguish between two (closely related) types of such assumptions:
\\[.2cm]
\noindent
\textbf{Closed World Assumption (CWA):} \textit{It is assumed that a priori -- in the formalization step of the problem -- an exhaustive, informative, and mutually exclusive list of all states relevant for the problem can be given and that the consequence that an act yields under each of  the states is clearly specifiable. Specifically, if a state is not contained in the list, it is assumed to be impossible to occur and is therefore irrelevant for the decision problem. }
\\[.2cm]
\noindent
Note that CWA originates from the theory of data bases \citep{reiter1978closed}, where it (essentially) states that what is currently not known about a system is assumed to be false. Beyond its relevance for classical decision theory (see \citet[Chapter 12]{gilboa2009theory} for a nice discussion), it is also judged to play a crucial role in more advanced areas like Markov Decision Processes (e.g., \cite{boutilier1995planning,DEARDEN1997219}) and, therefore, also appears to be of immense practical relevance for, e.g., modern machine learning.
\\[.2cm]
\noindent
\textbf{Small World Assumption (SWA):} \textit{It is assumed that there is a natural granularity of the states of the world for the considered decision problem. Any finer or coarser partition of the states of the world is judged to be accepted as unnatural. Specifically, no matter how complex the true processes driving the dynamics of the system under consideration, it is assumed that these dynamics can be precisely enough captured by a moderately large list of states of the world with an intuitive interpretation.}
\\[.2cm]
\noindent
Note that SWA is also closely related (but not identical) to the problem of \textit{act-state dependence}: In Savage's original model \citep{foundstat} for decision making under uncertainty (which, in essence, coincides with the one recalled in Section~\ref{classical1}), the acts are assumed (by construction) to not influence the availability and the probability of the states of nature. This, too, is an (indirect) assumption about the ease with which the world can be described in a way that resolves all uncertainty and, therefore, can be considered a sub-assumption of SWA.\footnote{This point of view is backed by the fact that the (presumably) most standard approach to circumvent act-state dependence is to extend the state space from $S$ to $A^S$, i.e., the set of all functions from $S$ to $A$ \citep{Gibbard1981}. While, in this way, it is ensured that all act-state dependence disappears from the model, as the extended state space definition directly accounts for potential dependencies, dealing with the extended state space enforces dealing with a state space of unnatural granularity, thereby enforcing violations of SWA.} The challenges arising from act-state dependence in decision problems have, prominently triggered by Newcomb's paradox \citep{Nozick1969}, established an entire subfield of modern decision theory (e.g., \citet{karni,karni1,ssk1,ssk2,jean,10.1007/978-3-031-08971-8_59}) with application ranging from insurance science (e.g.,~\citet{dedonder}) to portfolio selection (e.g., \citet{wei}).
\\[.2cm]
\noindent
But why can these assumptions be problematic? At first glance, it seems obvious to argue that both assumptions, CWA and SWA, are always trivially satisfiable: In the case of CWA, an artificial state $s^*$ could always be introduced, which occurs exactly when none of the clearly specifiable states occurs. However, closer inspection  
 clarifies that CWA also requires that under each state the consequences of the actions must be clearly specifiable. This is obviously less easily satisfied: What consequence should one assign to an artificial state $s^*$ that cannot even be precisely described? In this light, CWA seems a very strong assumption after all.
\\[.2cm]
\noindent
In the case of SWA one could try to argue that a choice function whose choice sets depend on the specific framing of the decision problem should not be chosen anyway, and thus invariance under the granularity of the set of states of the world should serve as a quality measure for choice functions rather than a criticism of SWA. However, the requirement of perfect invariance under the granularity of the state space (i.e., the way the states are represented as a set) quickly leads to trivial and non-informative decision criteria: Most often the only choice function satisfying this kind of invariance property will be the identity map, which is obviously useless for any meaningful analysis. Thus, invariance under the granularity of the state space is presumably only desirable at first sight. This point of view is certainly backed by the observation that even the choice function based on expected utility does not satisfy this requirement. Finally, it is important to note that the two assumptions -- CWA and SWA -- are by no means independent of each other: if an artificial state $s^*$ is introduced as described above, it is probably only in extremely rare cases possible to make statements about the naturalness of its granularity. In other words: often, the violation of one of the assumptions can only be prevented by deliberately violating the other assumption.
\begin{bsp}
Consider again the situation of our running Example \ref{mot_ex}. It is easy to see here that neither CWA nor SWA are sensibly acceptable. Since we only have exemplary consequences of certain decisions (the different amounts of monetary gains throughout the month) at our disposal, but no insight into the underlying dynamics, it is difficult in this situation to specify even an approximation of a state space. In such a situation, we clearly leave the scope of application of classical decision-theoretical methods and therefore need alternative tools to analyze the existing information (i.e. the observed consequences of action). Any serious alternative theory should, on the one hand, recognize the empirical nature of the problem but, on the other hand, also provide guarantees of optimality that go beyond the merely empirical.
\end{bsp}
%
%
\section{Empirical Decision Problems}\label{sec:edp}
\begin{figure}[h]
    \centering
    \includegraphics[width=.6\linewidth]{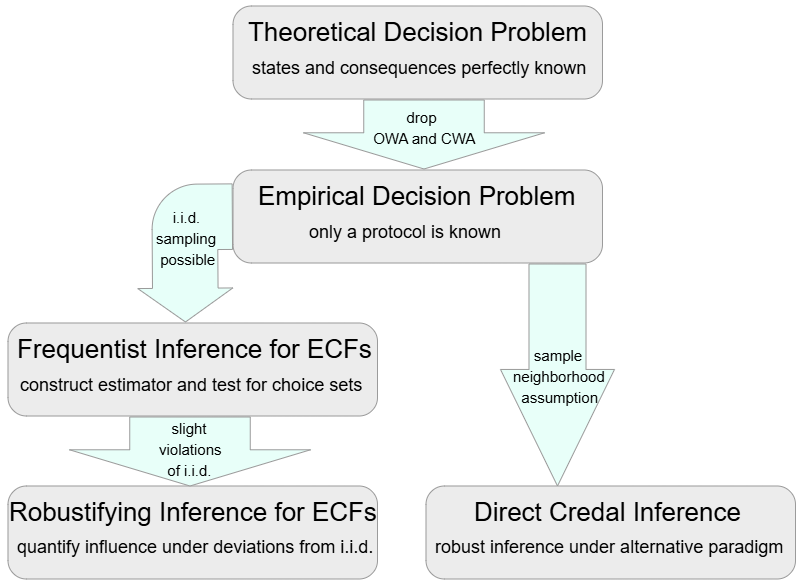}
    \caption{Overview on the concepts discussed in the present paper and how they interconnect with each other. At this point in time, we (essentially) only discussed the very top node of the diagram and will hence now start discussing the paper's core contribution.}
    \label{fig:summary}
\end{figure}

\noindent We now come to the main object of investigation of this paper: \textit{empirical decision problems (EDPs)}. The main motivation here is to build a decision theory exclusively on \textit{observable} quantities and thus to completely avoid strong assumptions like SWA and CWA and, with them, the necessity to define a state space or a set of consequences at all. At the same time, our newly built theory should be as close as possible to the classical decision-theoretic framework recalled in Section~\ref{known_states}, in this way, still retaining its technical and philosophical maturity. However, the sparseness of assumptions has to come at a price: Building solely on observables makes the statements about e.g. optimality of decisions purely empirical ones, not directly generalizing to the underlying true decision problem (if there is one). This demands to back up these statements with generalizable guarantees. \\[.2cm]
\noindent To this end, we propose three different approaches: In Section \ref{estimation}, we show under which conditions consistent estimators can be constructed for the choice sets of special classes of choice functions in an EDP. In Section~\ref{testing}, we derive statistical tests based solely on observed consequences for checking whether an act lies in the choice set of some choice function of interest. In Section~\ref{direct_inference}, we propose a \textit{direct credal inference} approach for robustifying empirically obtained choice functions by using credal sets for describing contaminated sampling models. A schematic overview of the paper so far and the concepts still to be discussed can be seen in Figure \ref{fig:summary}.
\subsection{The Formal Structure of an EDP}
%
\noindent
Let $\mathbb{A}=\{a_1, \dots , a_n\}$ be a \textit{finite} set.  The elements of $\mathbb{A}$ are called \textit{action descriptions} and refer to decisions within a specific context that the decision maker under consideration has already followed at least once in the past. Importantly, note that describing $a \in \mathbb{A}$ does neither require specifying a state space $S$ nor a consequence space $C$: It is a pure description of the action the decision maker is taking, for instance \textit{"choose the blue card to clock out"}. This is in stark contrast to the acts (see Section~\ref{classical1}) in the classical theory, which are functions from the state to the consequence space and therefore require both these sets to be known. In contrast, the idea of action descriptions is to -- in a first step -- separate the potential decision from the variable (or act) associated with it, thereby obtaining a somewhat more naive and natural description of the decision-making situation: \textit{People are able to verbalize courses of action without having to talk about mathematical objects.}
\\[.2cm]
\noindent
However, for each $a \in \mathbb{A}$, we \textit{do} assume there exists an associated \textit{act} $X_a:S_a \to C_a$ expressing the \textit{unknown} consequence $c \in C_a$ of following the action description $a$ as a function defined on a \textit{latent} state space $S_a$. We call the set $\mathbb{A}$ \textit{context consistent}, whenever there exists sets $C$ and $S$ such that $C_a=C$ and $S_a=S$ for all $a \in \mathbb{A}$, i.e., if the consequences of the action descriptions in $\mathbb{A}$ can be analyzed within the same context. Note that, while both sets $S$ and $C$ are assumed to be unknown to the decision maker, there is a major difference in their interpretation: While $S$ is assumed to be a latent space, i.e., to consist of \textit{unobservable} elements, the elements of $C$ are indeed observable: Just follow some action description and observe what happens!\footnote{It should be noted that for the inferential guarantees discussed in later Sections~\ref{estimation} and~\ref{testing} of the paper to work, some structural assumptions must also be made in addition to the mere observability of $C$.}
\\[.2cm]
\noindent
Having said this, let us move on by defining the necessary concepts: For a context-consistent set of action descriptions $\mathbb{A}$, we build our model based on \textit{act-consequence pairs}, i.e. observed pairs of the form $(a,c) \in \mathbb{A}\times C$ interpreted as: \textit{Following description $a$ leads to consequence $c$} or, to use our running example, \textit{clocking out with the blue card lead to a 5€ bonus on the canteen card}.  Let $\mathbf{z}:=(z_i)_{i \in \underline{n}}\subseteq \mathbb{N}$, where $z_i$ models the number of times that $a_i$ was already performed in the past and denote $\underline{n} = \{1, \dots, n\}$. A \textit{$\mathbf{z}$-protocol} $\mathscr{P}_{\mathbf{z}}$  is a finite family $((q_j,p_j))_{j \in \underline{N}}$ of act-consequence pairs, where $N=\sum_{i=1}^{n}z_i$, such that, for all $i \in \underline{n}$, it holds $|\{j \in \underline{N}: q_j=a_i\}|=z_i$.  For instance, if $z_3=8$, this means that action description $a_3$ was followed exactly eight times in the available protocol. Importantly, in general, the consequences associated with these different trials of $a_3$ will not be constant, but rather varying. This is due to the fact that $a_3$ is associated with a \textit{random} variable $X_{a_3}$ on a latent set of states of the world. We can now give the following central definition.
%
\begin{defi}\label{def:sdp}
Let $\mathbb{A}=\{a_1, \dots , a_n\}$ be a context-consistent set of action descriptions and let $\mathscr{P}_{\mathbf{z}}$ be a $\mathbf{z}$-protocol for $\mathbb{A}$. The pair $(\mathbb{A},\mathscr{P}_{\mathbf{z}})$ is called \textbf{empirical decision problem (EDP)}.
\end{defi}
Similar to classical decision problems, the main goal in EDPs is to define an \textit{empirical choice function} that returns a set of action descriptions for any sub-protocol inserted into it. Depending on the quality of the available information, this set is then interpreted as a set of optimal or non-rejectable action descriptions \textit{given the data} (i.e. the protocol), similar to the distinction between the weak and the strong view on classical choice sets given in Section \ref{known_states}. It is important to emphasize the empirical nature of EDPs: Two different protocols $\mathscr{P}_{\mathbf{z}}$ and $\mathscr{P}_{\mathbf{z}'}$ will, in general, lead to different EDPs and therefore empirical choice functions even if they are based on the same context-consistent decision problem $\{X_a: a \in \mathbb{A}\}$. Let us now put this more formally:  For every $M\subseteq \underline{n}$, denote by $\mathscr{P}_{\mathbf{z}}(M)$ that sub--family of $\mathscr{P}_{\mathbf{z}}$ containing only those act-consequence pairs $(a_i,c)$ satisfying $i \in M$, and set $S(\mathscr{P}_{\mathbf{z}}):=\{\mathscr{P}_{\mathbf{z}}(M): M \subseteq \underline{n}\}$. Each element $\mathscr{P}_{\mathbf{z}}(M)$ shall be called \textit{sub-protocol}. 
\begin{defi}\label{def:ecf}
Let $(\mathbb{A},\mathscr{P}_{\mathbf{z}})$ be an EDP. An \textbf{empirical choice function (ECF)} is a function 
$$\hat{\text{ch}}_{\mathbf{z}}: S(\mathscr{P}_{\mathbf{z}})\to 2^{\mathbb{A}}$$ 
satisfying $\hat{\text{ch}}_{\mathbf{z}}(\mathscr{P}_{\mathbf{z}}(M)) \subseteq \{a_i: i \in M\}$ for all $M \subseteq \underline{n}$.    
\end{defi}
We end this section by continuing our running example for llustration.
\begin{bsp}
Let us now assume that the employee used each of the cards exactly five times to clock out during the last month. The resulting protocol containing the corresponding amounts of money is shown in Table~\ref{ex3}.
\begin{table}[ht]
\centering
\caption{Trial Results by Color (Values in €)}\label{table_running_example}
\begin{tabular}{
  >{\bfseries}l 
  *{5}{>{\centering\arraybackslash}S[table-format=2.0, table-number-alignment=center]}
}%
\toprule
Color     & {Trial 1} & {Trial 2} & {Trial 3} & {Trial 4} & {Trial 5} \\
\midrule
Red       & 3€  & 1€  & 6€  & 2€  & 5€ \\
Blue      & 1€  & 7€  & 2€  & 1€  & 3€ \\
Green     & 5€  & 2€  & 1€  & 7€  & 0€ \\
Yellow    & 2€  & 1€  & 3€  & 0€  & 4€ \\
Black     & 1€  & 2€  & 1€  & 3€  & 0€ \\
\bottomrule
\end{tabular}
\label{ex3}
\end{table}
One way to construct an ECF for this example is to use an \textbf{empirical variant} of the EU choice function from Equation~\eqref{EU}: For each colored card, we compute the empirical expectation based on the corresponding empirical probability measure. Then, we collect all colors leading to the maximal empirical expectation. In this example, out of all cards, this ECF selects only \textbf{Red},  as $\tfrac{17}{5}$ is the unique maximum value of the expectations. Another way to construct an ECF is to use an \textbf{empirical variant} of the FSD choice function from Equation~\eqref{FSD}. To this end, we calculate the infimal empirical expectation difference with respect to the corresponding mass for all pairs of colors, where the infimum is taken over all monotonic transformations of the set $\{0,\dots , 7\}$. Finally, in order to compose the empirical choice set, we collect all colors where this difference is always at least $0$ for each competing color. In this example, the ECF based on empirical FSD collects $\{\textbf{Red}, \textbf{Blue}, \textbf{Green}\}.$\footnote{In particular, all acts in the group $\{\textbf{Yellow}, \textbf{Black}\}$  are empirically dominated by all acts in the group \{\textbf{Red}, \textbf{Blue}, \textbf{Green}\} and within both groups there is empirical incomparability. This can be immediately seen by increasingly ordering the values within each row of Table~\ref{table_running_example} and by recognizing that each trial corresponds to the same empirical probability of $1/5$ and that 
e.g.~\textbf{Red} dominates \textbf{Black} because the values in each column of the row-wise ordered table is now larger for \textbf{Red}, compared to \textbf{Black}.} Importantly, note the difference to classical decision problems: These optimal empirical choice sets are \textbf{dependent} on the observed protocol and, as such, may change after further experimentation with choosing the cards.
\end{bsp}
\subsection{Limitations and Opportunities of Empirical Decision Theory} \label{limitations}
\noindent The usefulness of the analysis of an EDP stands and falls with the interpretation of the protocol it contains. Put more specifically, if the protocol for each concrete action description $a_i$ can be understood as a random sample from the distribution of $X_{a_i}$, i.e., the act underlying the action description, then EDPs can be approached using specialized techniques of (non-parametric robust) frequentist statistics. This will be the main route we will follow in the next few sections of the present article. In Section \ref{cont_prot}, we will even see that some of the methods developed can be extended to cases where the classical statistical assumptions are slightly violated.
\\[.2cm]
Nevertheless, a clear limitation of the scope of our theory is already apparent here: To ensure that EDPs can be analyzed using frequentist statistical methods, even if only potentially, we assume that each of the generations of the states of the world is happening according to a (generally unknown) probability distribution—i.e., it stems from the independent repetitions of the same random experiment. This assumption can, and will be, weakened by considering contamination models. Even more severe violations, however, would allow almost anything to happen, and analyses based on protocols or EDPs in general could become highly misleading. An especially severe example of such a situation is illustrated in the following continuation of our running problem.
\begin{bsp}
Let's assume that we are back in the situation of our running example, but before our employee started experimenting with the different colored cards (i.e., no protocol has been collected so far). We further assume that the true underlying system has only three states -- $s_1,s_2,s_3$ -- and functions as follows:
\begin{table}[ht]
\centering
\caption{Groundtruth Decision Problem (Values in €)}
\begin{tabular}{
  >{\bfseries}l 
  *{3}{>{\centering\arraybackslash}S[table-format=2.0, table-number-alignment=center]}
}
\toprule
Color     & {State $s_1$} & {State $s_2$} & {State $s_3$} \\
\midrule
Red       & 4€  & 1€  & 1€   \\
Blue      & 6€  & 3€  & 2€   \\
Green     & 5€  & 3€  & 3€   \\
Yellow    & 10€  & 2€  & 2€   \\
Black     & 8€  & 2€  & 3€   \\
\bottomrule
\end{tabular}
\end{table}
As before, our employee is unaware of this underlying decision-making problem, and they begin to experiment. Unlike before, however, we now explicitly assume that the states of the system (during the experimental phase) are not random, but are generated according to a simple deterministic strategy: Put the system to state $s_1$ if \textbf{Red} is chosen, and to $s_3$ otherwise. When the experimental phase is complete, however, the system will be switched to random generation of its states (according to some distribution $\pi$). If the employee starts experimenting, no matter how many trials, a very simple structured protocol emerges: Every example that starts with the red card is assigned a profit of $4$€, and every example that starts with a card of a different color is assigned a profit of strictly less than $4$€. It is very easy to see that both ECFs -- the one based on \eqref{EU} (with utilities equaling profits) and the one based on \eqref{FSD} -- would uniquely recommend choosing the red card for any protocol of this structure. And this despite the fact that, in terms of decision theory, choosing the red card is strictly inadmissible in the underlying decision problem: every other card gives strictly more profit for no matter what state the system ends up in.
\end{bsp}
The above discussion makes it abundantly clear that the analysis of EDPs is usually only useful if the underlying system does not produce its states by a simple deterministic structure, but rather according to a (generally unknown) mechanism that is / can be taken as random. This is the only way to ensure that the consequences recorded in the protocol actually approximate the realities of the underlying decision-making problem. In the following, we will therefore always assume that we are dealing with well-behaving situations in exactly this sense. But before, we conclude this section by exemplarily discussing two situations that underline the possibilities that the analysis of EDPs offers from an illustrative perspective.
\begin{bsp}
Suppose our employee learns that all acts underlying the choices of the different clock-out cards follow a specific $Bin(10,p)$ distribution. For the sake of illustration, we discuss two different scenarios: First, assume the parameters associated with the different colors are given as $(p_{red}, \dots ,p_{black})=(0.25,0.2,0.22,0.22,0.21)$. Figure~\ref{fig:EUex} shows the development of the choice sets of the empirical EU choice function along increasing protocol size (where the consequences under the different action descriptions are sampled \textit{i.i.d.} and per round we sample one consequence under each action description). We observe that the choice sets stabilize at the singleton set containing only the red card rather quickly (which is also the true EU maximizer in the underlying population decision problem). Second, assume the parameters associated with the different colors are given as $(p_{red}, \dots ,p_{black})=(0.32,0.32,0.45,0.8,0.8)$. Figure~\ref{fig:FSDex} shows the evolution of the choice sets of empirical FSD along increasing protocol size (still assuming \textit{i.i.d.}~sampling). In contrast to the simulation under EU, we can observe here that the empirical choice sets do \textbf{not} appear to stabilize. That is, even though we know that the true choice set of the population-based FSD-choice function is given by $\{X_{black}, X_{yellow}\}$  (this set being the true choice set follows from the fact that binomial distributions with fixed size are FSD-ordered according to the size of their respective parameter $p$), we repeatedly receive supersets of this set also after a relatively high number of simulation rounds. The reason for this will be explained in our later Theorem~\ref{theorem_consistent_approximator}: In order to guarantee that the empirical choice sets also stabilize under FSD, the “naive” empirical variant of the FSD choice function used here must be replaced by a regularized version.
    \begin{figure}
    \centering
\includegraphics[scale=.57]{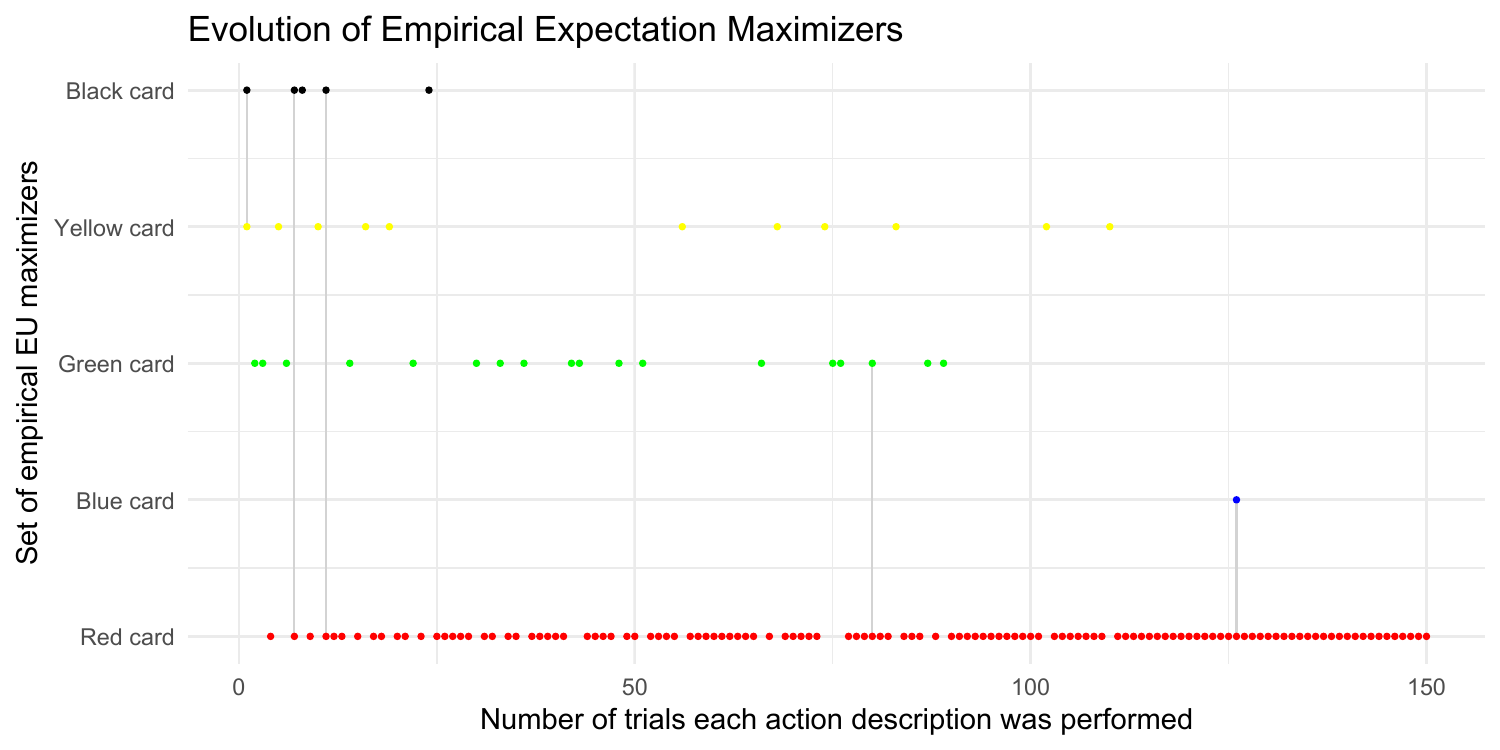}
    \caption{Choice sets of empirical EU along increasing protocol size. At each stage, the choice sets consists of all colored points that are connected by a vertical gray line.}
    \label{fig:EUex}
\end{figure}
    \begin{figure}
    \centering
\includegraphics[scale=.57]{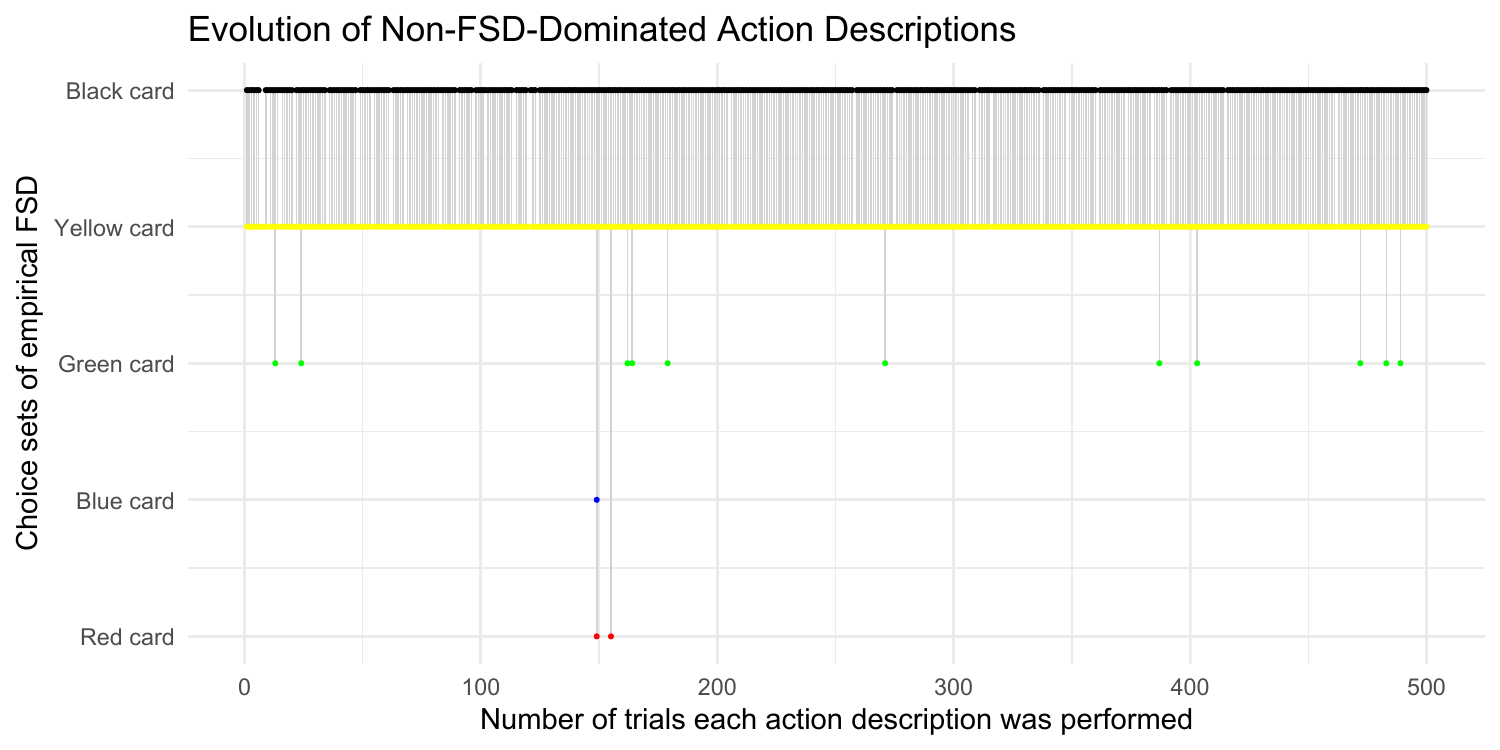}
    \caption{Choice sets of empirical FSD along increasing protocol size. At each stage, the choice sets consists of all colored points that are connected by a vertical gray line.}
    \label{fig:FSDex}
\end{figure}
\end{bsp}
This example suggests two important insights: First, that ECFs are applicable whenever a protocol is available, without making any significant assumptions about the nature of the underlying decision problem. This is a clear advantage compared to the traditional approach of decision theory and thus clearly demonstrates the possibilities of empirical decision theory. Secondly, that the empirical choice sets of certain choice functions can stabilize under the assumption of identical and independently distributed consequence observations in the true choice sets of the underlying population-based choice function. We will now formally substantiate this intuition.
\section{Statistical Estimation in EDPs} \label{estimation}
\noindent We now turn to the question of estimating true, i.e., population-based choice sets from empirical ones. Recall that an empirical choice function assigns sub-protocols of act-consequence pairs to subsets of the action descriptions which are relevant in these sub-protocols, i.e., present in at least one of the examples contained in the protocol. It thereby formalizes the process of selecting the most appropriate action descriptions from the inserted subprotocol. Intuitively, the selections proposed by an empirical choice function should get more and more appropriate as the \textit{protocol size} $\mathbf{\underline{z}}:=\min \{z_i : i \in \underline{n}\}$ increases. But how can we formalize this intuition? To answer this question, we first need to equip a given empirical choice function $\hat{\text{ch}}_{\mathbf{z}}$ with some additional interpretation, namely as an attempt to approximate an underlying \textit{population} choice function $\text{ch}:2^{X_{\mathbb{A}}} \to 2^{X_{\mathbb{A}}}$.
\\[.2cm]
\noindent The next step towards an answer concerns the protocol itself: We interpret each sub-protocol $\mathscr{P}_{\mathbf{z}}(\{i\})$, where $i \in \underline{n}$, as a sample of size $z_i$ from the distribution of the \textit{latent} variable $X_{a_i}$. Note that for this we implicitly assume that there exists some (unknown) probability $\pi$ producing the latent states of the world $s \in S$, making $\mathscr{P}_{\mathbf{z}}(\{i\})$ a sample from $\pi_{X_{a_i}}$. We can now formalize the intuition of an empirical choice function getting more and more appropriate as $\mathbf{\underline{z}}$ increases.
\begin{defi}\label{def:consistent-approx}
Let $(\mathbb{A},\mathscr{P}_{\mathbf{z}})$ be an empirical decision problem. An ECF $\hat{\text{ch}}_{\mathbf{z}}$ is called \textbf{consistent approximator} of a choice function $\text{ch}:2^{X_{\mathbb{A}}} \to 2^{X_{\mathbb{A}}}$, if  for all $ M \subseteq \underline{n}$ we have
 \begin{equation}
         \pi \Biggl(\Bigr\{ \lim_{\mathbf{\underline{z}}\to \infty} X_{\hat{\text{ch}}_{\mathbf{z}}(\mathscr{P}_{\mathbf{z}}(M))}=\text{ch}(\{X_{a_i}: i \in M\}) \Bigl\} \Biggr)=1,
    \end{equation}
    where $X_{\mathbb{A}'}:=\{X_a: a \in \mathbb{A}'\}$ for all $\mathbb{A}'\subseteq \mathbb{A}$ and set convergence is defined via the trivial metric.
\end{defi}
Let's have a careful look at an (implicit) implication of the consistency criterion just defined: When we assume $\mathbf{\underline{z}}\to \infty$, we implicitly agree that each of the available action descriptions in $\mathbb{A}$ can (at least potentially) be followed arbitrarily often. This assumption rules out agents applying a strategy along the lines of \textit{trying each description once and then going with the one having turned out best for them when played first}. Thus, we consider agents that \textit{actually believe} in the uncertain and non-deterministic nature of the consequences of their actions. Note, while articulated in a different context, this is strongly related to the concept of \textit{experimentation} in revealed preference theory, which was recently investigated and axiomatically characterized in \citet{ok2022indifference}.\\[.2cm]
\noindent But under what conditions can an ECF fulfill the consistency criterion just introduced? Intuitively speaking, it must be sensitive to the distribution of the variables to which it is applied. Sensitivity has two meanings here: firstly, the construction of the choice sets should depend on the distribution of the involved variables. Secondly, small changes in the distribution of the involved variables should only lead to small changes in the resulting choice set, as a (still to be defined) continuity in the variable distributions. If these two conditions do not apply, it is difficult to expect that the choice sets of the ECF will converge to those of the population CF as the empirical variable distributions converge to the true ones. We start by giving a definition capturing the intuition of a choice function being sensitive to the distribution of the variables.
\begin{defi}\label{distchoic}
For $k \in \mathbb{N}$ and $\mathcal{D}\subseteq \mathcal{G}$, define $$\binom{\mathcal{D}}{ k}:=\{A \in  2^\mathcal{D}\mid |A|=k\} ~~~~~~~~~~~~\text{ and } ~~~~~~~~~~~~~ \mathcal{D}_k:=\bigcup\limits_{i=1}^k \binom{\mathcal{D}}{ k}.$$ A choice function $ch$ is said to be a \textbf{$k$-ary choice function} if there exists a criterion function $cr: \mathcal{G}\times \mathcal{G}_{k-1} \longrightarrow  \mathbb{R}^2$ and a threshold function $c: 2^\mathcal{G}:\longrightarrow \mathbb{R}^{k-1} \times \mathbb{R}^{k-1}$ such that~$ch$~can be represented by locally comparing an act $A\in\mathcal{D}$ to up to $k-1$ further acts in the choice set $\mathcal{D}$ via:
\begin{align}\label{criterion_characterization}    
A \notin ch(\mathcal{D}) \iff \exists \mathcal{B}\in \mathcal{D}_{ k-1} : cr((A,\mathcal{B}))_1\geq (c(\mathcal{D})_1)_{|\mathcal{B}|} ,\; cr((A,\mathcal{B}))_2 > (c(\mathcal{D})_2)_{|\mathcal{B}|}.
\end{align}
For the special case $k=1$, by convention we set the domain of $cr$ to $\mathcal{G}$, the codomain of $c$ to $\mathbb{R} \times \mathbb{R}$ and $A \notin ch(\mathcal{D}) \iff  cr(A)_1\geq c(\mathcal{D})_1\;,\; cr(A)_2 > c(\mathcal{D})_2$ . Furthermore, if $cr$ depends on the acts only via their distributions, we call $ch$ a \textbf{distributional $k$-ary choice function}.
\end{defi}
    This definition appears to be a little bit involved. This is because it intends to capture arbitrary choice functions (with the set of all choice sets of  cardinality smaller than or equal to $k$ as a domain.) In the sequel, we will focus on binary distributional choice functions and a threshold function that is constantly zero. (And we abuse notation by writing $cr(A,B)$ instead of $cr(A,\{B\})$.) A prototypical example would be an EU-based choice function $ch_\mathcal{F}$ given by the criterion function $$cr(X_{a_i},X_{a_j}) =\left(-\inf \limits_{f \in \mathcal{F}} \mathbb{E}_\pi\left(f\circ X_{a_i} -f \circ X_{a_j}\right) , \inf \limits_{f \in \mathcal{F}} \mathbb{E}_\pi\left(f\circ X_{a_j} -f \circ X_{a_i}\right)\right).$$ But note that not all binary choice functions are of exactly this form, compare Remark~\ref{remark_sufficient_for_consistency}.



\begin{theorem}\label{theorem_consistent_approximator}
Let $ch$ be such that there exists a class $\mathcal{F}$ of uniformly bounded and measurable functions such that $cr$ is uniformly continuous  w.r.t. the product pseudometric of the pseudometric $\mathfrak{d}(X,Y):=\sup\nolimits_{f \in \mathcal{F} } | \mathbb{E}(  f \circ X   - f\circ Y)|$. Note that this pseudometric is only dependent on the involved (marginal) distributions of $X$ and $Y$. Let furthermore $\hat{ch}_\mathbf{{z}}$ be defined as the corresponding regularized empirical analogue for $ch$ in the form the following criterion: for $i \in M$ define 
\begin{align}
 a_i \notin \hat{ch}_\mathbf{{z}}(\mathscr{P}_{\mathbf{z}}(M)) & \iff \exists a_j \neq a_i: cr(\hat{X}_{a_i}, \hat{X}_{a_j})_{1 }\ >0\; , \; cr(\hat{X}_{a_i}, \hat{X}_{a_j})_{2 } +4\cdot \delta(\varepsilon_{\underline{z}}) \geq 0,\label{definition-empirical_choice_function}
\end{align}
\noindent where $\hat{X}_{a_i}$ and $\hat{X}_{a_j}$ are the random variables defined via the observed outcomes in the protocol, equipped with their empirical distributions and all associated characteristics like expectation etc. and $\varepsilon_{\mathbf{\underline{z}}} \in \Theta(1/\sqrt[4]{\mathbf{\underline{z}}})$ and $\delta$ is the isotone function defined by $$\delta: \mathbb{R}_{\geq 0}\longrightarrow \mathbb{R}_{\geq 0} \cup \{\infty\}: \varepsilon \mapsto \sup\limits_{\underset{\mathfrak{d}(X,\tilde{X})\leq \varepsilon, \mathfrak{d}(Y,\tilde{Y})\leq \varepsilon}{X,Y,\tilde{X},\tilde{Y},}} | cr_2(X,Y)-cr_2(\tilde{X},\tilde{Y})|.$$ Note that because $\mathfrak{d}$ is uniformly continuous, for $\varepsilon$ small enough, $\delta(\varepsilon)$ will always be finite and $\lim\nolimits_{\varepsilon \rightarrow 0}\delta(\varepsilon)=0$. If the family $\mathfrak{C}:=\{ \{c \in C \mid f(c) \geq \alpha\}\mid f \in \mathcal{F}, \alpha \in \mathbb{R}\}$  has a finite VC dimension and if furthermore $cr$ is antisymmetric (in the sense of $cr(X_{a_i},X_{a_j})_1 \geq0 , cr(X_{a_j},X_{a_i})_2\geq 0 \Longrightarrow a_i = a_j$), 
    then $\hat{ch}_\mathbf{{z}}$ is a consistent approximator of $ch$, given the protocols are sampled i.i.d.\footnote{Here, i.i.d. means that for fixed action description $a_i$ the observed outcomes are independently and identically distributed. The observations for different action descriptions need not to be independent. This is a difference to Assumption~\ref{as4} needed in Theorem~\ref{theo2}.}
\end{theorem}

\noindent \textit{Proof.} See Appendix.\hfill $\square$
\begin{remark}\label{remark_sufficient_for_consistency}
    If $cr$ is defined in a way that characterizes a choice function $ch_\mathcal{F}$ via $$cr(X_{a_i},X_{a_j}) =\left(-\inf \limits_{f \in \mathcal{F}} \mathbb{E}_\pi\left(f\circ X_{a_i}) -f \circ X_{a_j}\right) , \inf \limits_{f \in \mathcal{F}} \mathbb{E}_\pi\left(f\circ X_{a_j} -f \circ X_{a_i}\right)\right),$$ then the assumption of a finite VC dimension of $\mathfrak{C}$ in Theorem~\ref{theorem_consistent_approximator} is e.g.~satisfied if:
    \begin{enumerate}
        \item $\mathcal{F}$ is finite
        \item $C$ is finite.
         \item $C \subseteq \mathbb{R}^d$ bounded and $\mathcal{F}$ is a family of linear functions (restricted to $C$).

        \item $C=C_1\times O_1\times O_2 \times \ldots \times O_k$ with $C_1\subseteq \mathbb{R}^1$ and $O_1,\ldots , O_k$ finite and $\mathcal{F}$ is a set of functions that are linear and isotone in the first component and isotone in the other components. 
        \item $C =\mathbb{R}^d$   and $cr$ characterizes the upper orthant stochastic order (cf., e.g., \cite[p.~140]{shaked_1994}
        ), ,i.e. $\mathcal{F} =\{\mathds{1}_{x \geq c }\mid c\in C\}$.
    \end{enumerate}
\noindent    Additionally, for $C=\mathbb{R}^1$, if $ch$ is defined as choosing all acts/protocols, that are nondominated w.r.t. the value at risk (or another quantile-based criterion that is uniformly continuosly dependent on a (set of) quantile(s)), then the assumption of Theorem~\ref{theorem_consistent_approximator} are also fulfilled under the additional assumption that the image laws $\pi_u$ have fixed support $supp=[a,b] \in \mathbb{R}^1$ and are absolutely continuously distributed with a Lebesgue-density that is uniformly bounded from below. (This ensures the uniform continuity of $cr$.) Note that such quantile-based choice functions appear to be not representable as EU-based choice functions in the sense of Definition~{\ref{EUbased}}.\footnote{To see this, take for example $C=[0,1]$ and $X,Y$ distributed with strictly increasing cumulative distribution functions on the whole interval $[0,1]$ and the choice function that chooses all acts with undominated median. Then, take $X$ and $Y$ with the same median and $Y$ strictly stochastically smaller than $X$ at every point except the median. Then, if $\mathcal{F}$ contains at least one istone function $f$ that is strictly isotone at least on some small interval (which is a not too restrictive demand), one would get $\mathbb{E}(f\circ Y) < \mathbb{E}(f\circ X)$ which would already be in contradiction to the fact that $X$ and $Y$ are indifferent w.r.t. the median order.}
    
\end{remark}
\noindent The assumption that $cr$ is antisymmetric is in fact a very restrictive assumption, as it excludes for example the case where $a_i \neq a_j$ but $\pi_{X_{a_i}}=\pi_{X_{a_j}}$. However, if $cr$ is \textit{mirror-symmetrical}, i.e., $cr(X,Y)_1=-cr(Y,X)_1$ and \textit{transitive}, i.e., $cr(X,Y)_1 >0, cr(Y,Z)_1 >0 \Longrightarrow cr(X,Z)_1>0$, then (under the assumptions of the theorem) the ECF asymptotically chooses exactly one action from a set of different actions that are equivalent in the sense of satisfying antisymmetry for $cr$ (where asymptotically means that almost surely there exists an index $z_0$ such that starting from $\mathbf{\underline{Z}} \geq z_0$ one of the actions is chosen (but not necessarily always the same for different $\mathbf{\underline{Z}} \geq z_0$)).
To avoid the assumption of antisymmetry, one could alternatively additionally regularize the first condition in (\ref{definition-empirical_choice_function}) from $cr(\hat{X}_{a_i}, \hat{X}_{a_j})_{1 }\ >0$ to e.g. $cr(\hat{X}_{a_i}, \hat{X}_{a_j})_{1 }\ > 4\cdot \delta(\varepsilon_{\underline{z}})$. Then, with  arguments  analogous to that of the proof of Theorem~\ref{theorem_consistent_approximator} one can show consistency of the 
modified ECF without assuming antisymmetry, however, with the need of a further regularization and particularly with the need of a concrete choice of the exact regularizer.

%
%
\section{Statistical Testing in EDPs} \label{testing}
\noindent We now turn to the question of \textbf{\textit{statistically testing}} whether a specific action description $a_i$ -- or more precisely its associated latent act $X_{a_i}$ -- \textit{significantly} lies in a choice set of some (population-based) choice function to a pre-defined error level $\alpha \in [0,1]$. In other words, we ask the question: Given only a protocol of pairs of action descriptions and consequences, \textit{is it possible to control the error probability of treating a choice set induced by an empirical choice function as if it was its population counterpart?} We will see that a positive answer to this question is possible, if a number of assumptions on a) the nature of the underlying choice function, b) the statistic used for testing, and c) the way the protocol is produced are satisfied.
\subsection{Testing with i.i.d.~Protocols}
\noindent Given an empirical decision problem $(\mathbb{A},\mathscr{P}_{\mathbf{z}})$ as in Definition~\ref{def:sdp}, a (population-based) choice function $\text{ch}:2^{X_{\mathbb{A}}} \to 2^{X_{\mathbb{A}}}$, a set $\mathcal{D} \subseteq X_{\mathbb{A}}$ of acts, and $i \in \underline{n}$ such that $X_{a_i} \in \mathcal{D}$, we are interested in constructing statistical tests for the pair of hypotheses
\begin{equation}
   H_0: X_{a_i} \notin ch(\mathcal{D}) ~~ \text{ vs. }~~ H_1: X_{a_i} \in ch(\mathcal{D}), 
\label{global}
\end{equation}
with a test statistic that \textit{only depends} on $(\mathbb{A},\mathscr{P}_{\mathbf{z}})$, i.e., the protocol of act consequence pairs the agent has experienced so far. In words, we want to find out whether there is enough evidence at a given probabilistic error level $\alpha$ to reject that a particular description is not in the true underlying choice set. Note that this testing problem is highly \textit{non-parametric} in the sense that no classical (parametric) statistical test seems to be applicable here. Although we essentially want to keep this non-parametric character to capture situations as general as possible, we will need to make a couple of assumptions to be able to design a meaningful test. The first one is:
\begin{as}\label{as1}
    The choice function is \textbf{induced by pairwise comparisons}: there is
a \textit{partial order} $\succsim$ on $X_{\mathbb{A}}$ with $ch(\mathcal{D}')= \{X: \nexists Y \in \mathcal{D}' \text{ s.t. } Y \succ X\}$ for all $\mathcal{D}' \subseteq X_{\mathbb{A}}$, with $\succ$ strict part of $\succsim$.
\end{as} 
\noindent In other words, the choice sets of $ch$ can be fully characterized as the not strictly dominated elements from $\mathcal{D}'$ with respect to the relation $\succsim$. Note that being induced by pairwise comparisons in the sense just defined is a strictly stronger requirement for a choice function than being a \textit{binary} choice function in the sense of Definition \ref{distchoic}: The set of binary choice functions corresponds exactly to the set of all choice functions that are induced as the not strictly dominated acts of \textit{arbitrary} binary relations. In contrast, Assumption~\ref{as1} refers to choice functions that are induced as the not strictly dominated elements of a \textit{partial order}, additionally ensuring that the inducer is transitive and antisymmetric. Both requirements are needed in our construction: First, note that under Assumption~\ref{as1} it is immediate that $H_0$ in (\ref{global}) is false, if for every $X_{a_j} \in \mathcal{D}\setminus\{X_{a_i}\}$ the hypothesis
\begin{equation}\label{composite}
    H_0^{ji}: X_{a_j} \succsim X_{a_i}
\end{equation}
is false. Here, we use that $\succsim$ is antisymmetric and that $X_{a_j} \neq X_{a_i}$ by assumption. This connection also translates to the error bounds for statistical tests: If we manage to find valid level $\alpha$ tests, $\alpha \in (0,1)$, for each of the pairs $(H_0^{ji},\neg H_0^{ji})$, we can construct a valid level-$\alpha$ test for the pair  $(H_0, H_1)$ by simply rejecting $H_0$ if, and only if, we reject every single $H_0^{ji}$. The validity of a so constructed test follows by observing
\begin{eqnarray*}
P_{H_0}(\text{reject }H_0) &=& P_{H_0}(\text{reject }H_0^{ji} \text{ for all } X_{a_j} \in \mathcal{D}\setminus\{X_{a_i}\})\\
& \leq & P_{H_0}(\text{reject }H_0^{j^*i})\\
&=&P_{{H_0}^{j^*i}}(\text{reject }H_0^{j^*i}) \leq \alpha ,
\end{eqnarray*}
where $X_{a_{j^*}}$ is a fixed but arbitrary element of $\mathcal{D}\setminus\{X_{a_i}\}$, i.e., by the observation that the type I error probability of tests for the original pair of hypotheses is bounded from above by the type I error probability of tests for any fixed pair of auxiliary hypotheses (which itself is bounded from above by $\alpha$ as we assumed these tests to be valid level-$\alpha$ tests). This shows that under Assumption~\ref{as1} we can \textit{decompose} our original test problem $(H_0, H_1)$ into $|\mathcal{D}|-1$ pairwise test problems $(H_0^{ji},\neg H_0^{ji})$. But how can adequate tests be constructed for these (presumably) easier problems? To accomplish this, we have to make some more assumptions:
%
\begin{as}\label{as4}
    For all $X_{a_j} \in \mathcal{D}$, the sub-protocol $\mathscr{P}_{\mathbf{z}}(\{j\})$ forms an \textbf{i.i.d.}~sample from $\pi_{X_{a_j}}$ and for all $X_{a_i}\neq X_{a_j} \in \mathcal{D}$ the random vectors $\mathscr{P}_{\mathbf{z}}(\{i\})$ and $\mathscr{P}_{\mathbf{z}}(\{j\})$ are stochastically independent. 
\end{as}
%
Under Assumptions~\ref{as1} to \ref{as4}, it is well-known that a \textbf{\textit{two-sample permutation test}}~(as described, e.g., in~\cite{pg2012}) can be used to construct a valid level-$\alpha$ test for each of the (auxiliary) pairwise hypothesis pairs
\begin{equation}\label{localaux}
    \tilde{H}_0^{ji}: \pi_{X_{a_j}}= \pi_{X_{a_i}}~~ \text{ vs. }~~ \tilde{H}_1^{ji}:\pi_{X_{a_j}} \neq \pi_{X_{a_i}}.
\end{equation}
Following an analogous decomposition argument as in the computations below Equation~\eqref{composite}, we can use this to construct a valid level-$\alpha$ test for the (auxiliary) global hypothesis pair
\begin{equation}\label{globalaux}
    \tilde{H}_0: \forall j\neq i :\pi_{X_{a_j}}= \pi_{X_{a_i}} ~~ \text{ vs. }~~ \tilde{H}_1:\exists j\neq i: \pi_{X_{a_j}} \neq  \pi_{X_{a_i}},
\end{equation}
by rejecting $\tilde{H}_0$ whenever we reject all individual permutation tests reject $\tilde{H}_0^{ji}$.
The \textbf{\textit{resampling scheme}} for $\tilde{H}_0^{ji}$ looks as follows: Start by pooling $\mathscr{P}_{\mathbf{z}}(\{i\})$ and $\mathscr{P}_{\mathbf{z}}(\{j\})$ to the combined sample 
$$W:= (w_1 , \dots , w_{\ell}):=(\mathscr{P}_{\mathbf{z}}(\{j\}),\mathscr{P}_{\mathbf{z}}(\{i\}))$$ 
Next, create a large number $N$ of resamples $((W_k^j,W_k^i))_{k \in \underline{N}}$, where each $W_k^j$ is a $|\mathscr{P}_{\mathbf{z}}(\{j\})|$-sample from $W$ drawn \textit{without replacement} and $W_k^i:=W \setminus W_k^j$ are the remaining data points from the pooled sample $W$. If we now denote by $T$ the test statistic and by $t_k^{ji}$ its realisation evaluated for the (artificial) sample $(W_k^j,W_k^i)$, we compare the value of the test statistic $t_0^{ji}$ computed based on the observed sample with the suitable quantiles of the empirical distribution of the resample statistics $(t_k^{ji})_{k \in \underline{N}}$ in order to derive a critical region for the test. Alternatively to a permutation test, one can also do a \textbf{\textit{bootstrap test}}. For a bootstrap test resampling is done with replacement instead and the test will typically only be asymptotically valid. \\[.2cm]
\noindent With the help of a series of permutation tests, we can thus construct a valid statistical test for hypothesis pair from \eqref{globalaux}. The question now is how this helps us in constructing a test for hypothesis pair \eqref{global}, which is actually of interest. To make this possible, we must make another assumption about the nature of the underlying choice function: it must be compatible with the null hypothesis of the permutation test. Only then performing a permutation test can potentially be informative for the question of actual interest. We add the following assumption to our list.
%
\begin{as}\label{as2}
 The choice function $ch$ is constructed in a way such that $H_0^{ji}$ \textit{\textbf{contains distributional equivalence}}, i.e.~the case where $\pi_{X_{a_i}}= \pi_{X_{a_j}}$.   
\end{as}
\noindent Assumption~\ref{as2} ensures that -- even if the null hypothesis $H_0^{ji}$ is compatible with many different distributions -- it is also compatible with the case where $X_{a_j}$ and $X_{a_i}$ follow the same distribution. While this assumption is obviously necessary in order to transfer the rejection regions of the permutation test to our test problem, it is unfortunately generally  not sufficient: If the interaction between the choice function used and the test statistic used is not restricted further, it may well happen that the null hypothesis induced by them is compatible with distributional equilivalence, but the test statistic induces overly optimistic rejection regions under distributional equivalence. To avoid this, it must be ensured that the test statistic assumes its worst-case distribution under this specific case.  However, how such an assumption can be mathematically expressed in concrete terms appears to depend heavily on the nature of the underlying choice function and the test statistic used. Here, we address only the specific case of an EU-based choice function $ch_{\mathcal{F}}$ in the sense of Definition~\ref{EUbased} and leave the elaboration for other classes of choice functions to future research.
%
\begin{as}\label{as3}
The test statistic $T$ attains its \textbf{\textit{least favorable distribution}} 
under $H_0^{ji}$ for distributional equivalence: 
for random variables $X,Y$ with $X \precsim Y$ (which corresponds to an arbitrary case of $H_0^{ji}$), we have (assuming negative values of $T$ point towards rejection) $$T(X^{(n)}, Y^{(m)}) \succsim_{FSD}  T(\tilde{X}^{(n)}, \tilde{Y}^{(m)}),$$ where $X^{(n)}=(X_1,\ldots,X_n) \stackrel{i.i.d.}{\sim} \pi_X, Y^{(m)}=(Y_1,\ldots,Y_m) \stackrel{i.i.d.}{\sim} \pi_Y$  and $\tilde{X}_1,\ldots , \tilde{X}_n, \tilde{Y}_1,\ldots , \tilde{Y}_m $ is a resample of the pooled sample $(X_1,\ldots,X_n,Y_1,\ldots,Y_m)$, without replacement for the permutation test or with replacement for the bootstrap test. 
If only $$\lim\nolimits_{m,n\rightarrow \infty} \lambda_{m,n}T(X^{(n)}, Y^{(m)}) \succsim_{FSD}  \lim\nolimits_{m,n\rightarrow \infty}\lambda_{m,n}T(\tilde{X}^{(n)}, \tilde{Y}^{(m)})$$ holds for some appropriately chosen normalization $\lambda_{m,n}$ that ensures non-degenerate limits, we say that Assumption \ref{as3} holds asymptotically.
Additionally, we assume that for $i,j \in\underline{n}$ we have $\lim\nolimits_{N\rightarrow \infty} z_i/z_j =\lambda_{ij} \in(0,1)$. Finally, we assume that the laws $\pi_{X_{a_i}}$ are non-degenerate in the sense that for every $i\neq j$ there exists some $f\in \mathcal{F}$ such that $\pi(f \circ X_{a_i} =f \circ X_{a_j}) <1$.  
\end{as}
\noindent 
Assumption~\ref{as3} provides a clear mathematical description of what it means to be a least favorable distribution under an EU-based choice function. Equipped with the generic Assumptions \ref{as1} to \ref{as4} and the EU-specific Assumption \ref{as3}, we can now finally arrive at the main result of this chapter, namely a complete description of a test for EU-based choice functions. We have:
\begin{theorem} \label{theo2}
 Let $(\mathbb{A},\mathscr{P}_{\mathbf{z}})$ with $\mathbb{A}=\{a_1, \dots , a_n\}$ be a context-consistent EDP with associated acts $\mathbb{X}_{\mathbb{A}}$ taking values in $C$. Assume $ch:=ch_{\mathcal{F}}$, let $\pi$ be a probability measure on $S$ and let $\mathcal{F}$ be a set of (uniformly) bounded and measurable functions $f:C \to \mathbb{R}$ such that $\succsim_{\mathcal{F}}$ is antisymmetric on $X_{\mathbb{A}}$. Fix $\alpha \in(0,1/2)$ and 
define the test statistic
$$T_{U,V}:C^\infty \times C^\infty \to \mathbb{R}~~,~~(c,c')\mapsto \inf\nolimits_{f \in \mathcal{F}} \Bigl\{\mathbb{E}_c(f \circ U) -\mathbb{E}_{c'}(f \circ V)\Bigr\},$$
 where $C^\infty:=\bigcup\nolimits_{k=1}^\infty C^k$ and where $\mathbb{E}_c$ and $\mathbb{E}_{c'}$ denote the expectations with respect to the empirical measures induced by the vectors $c$ and $c'$. Then we have:
 \begin{enumerate}[i)]
     \item Under Assumptions \ref{as1} and \ref{as4} %
  performing a level-$\alpha$ two-sample permutation test for $(\tilde{H}_0^{ji},\tilde{H}_1^{ji})$ from \eqref{localaux}
 based on $T_{X_{a_j},X_{a_i}}(\mathscr{P}_{\mathbf{z}}(\{j\}),\mathscr{P}_{\mathbf{z}}(\{i\}))$ for each $X_{a_j}\in \mathcal{D}\setminus\{X_{a_i}\}$ and rejecting 
whenever all those tests reject, gives an exactly valid level-$\alpha$ test for $(\tilde{H}_0,\tilde{H}_1)$ from \eqref{globalaux}.
\item If additionally Assumption~\ref{as2} and the assumptions of Theorem~\ref{theorem_consistent_approximator} are satisfied, the test  described in i) is also consistent for the directed alternative $$H^*_1:\forall j\neq i: X_{a_j} \nsuccsim_{\mathcal{F}}  X_{a_i}.$$
\item If $T$ is always non-positive\footnote{This is the case for example if $\mathcal{F}$ contains the function $f_0$ that is always zero, or if it contains a certain function $f$ and  all scaled functions $1/n \cdot f$ with  $n \in \mathbb{N}_+$.} and if $\mathcal{F}$ is a uniformly bounded Donsker class for all involved laws $\pi_{X_{a_i}}$ with $i \in\underline{n}$ (and therefore also all possible mixtures thereof), then if additionally Assumption \ref{as3} holds (exactly or asymptotically), the permutation test described in $i)$ is also asymptotically valid for the hypothesis pair of actual interest
\begin{equation*}
   H_0: X_{a_i} \notin ch_{\mathcal{F}}(\mathcal{D}) ~~ \text{ vs. }~~ H_1: X_{a_i} \in ch_{\mathcal{F}}(\mathcal{D}) 
\end{equation*}
\item Statements $i)-iii)$ are analogously asymptotically valid for a bootstrap test (where, compared to the permutation resampling scheme, the subsampling is done with instead of without replacement).~Particularly, under the given assumptions the limiting processes of the bootstrapped and the permuted version of the original stochastic process $(\hat{\mathbb{E}}(f \circ U) - \hat{\mathbb{E}}(f\circ V))_{f\in\mathcal{F}}$ are equal in distribution.
Therefore, the same holds for the associated test statistics.

\end{enumerate}
\end{theorem}
\noindent\textit{Proof.} See Appendix. \hfill $\square$\\[.2cm]
    \noindent The Donsker class property in iii) is satisfied for example if $\mathcal{F}$ is uniformly bounded and has a finite subgraph VC dimension. One concrete example is FSD, which is characterized by the class $\mathcal{I}_\succsim$ of measurable isotone indicator functions in the case that $\succsim$ has finite width, which guarantees a finite VC dimension, cf., \citet[Proposition~2, p.~31]{sja2017}. Note that both the class $\mathcal{I}_\succsim$ and the class $\mathcal{U}_\succsim$ generate the same preorder and therefore the same choice function. Additionally, the characterizing test statistics $T_{U,V}$ are also  identical, because $$\inf\nolimits_{f \in \mathcal{U}_\succsim}\left\{ E_c(f\circ U) - E_{c'}(f\circ V)\right\} = \inf\nolimits_{f \in \mathcal{I}_\succsim}\left\{ E_c(f\circ U) - E_{c'}(f\circ V)\right\},$$ see, e.g., \citet[p.~9]{sja2017}. For FSD, also Assumption~\ref{as3} is satisfied exactly. More generally, if $\mathscr{S}\subseteq 2^C$ is a complete ring of sets\footnote{A complete ring of sets is a family of sets that is closed under arbitrary intersections and unions (including the empty intersection (i.e., $C$) and the empty union (i.e., $\emptyset$)).}, then Assumption \ref{as3} is fulfilled exactly for the class  $\mathcal{F}:=\{1_S\mid S\in\mathscr{S}\}$ (or the convex cone generated by $\mathcal{F}$). 
    Finally, under the Donsker class assumption of $iii)$, Assumption~ \ref{as3} is also satisfied asymptotically, see the proof of Theorem~\ref{theo2}. Therefore, this assumption is actually not explicitly needed in $iii)$. However, we added this assumption explicitly, because in certain situations, like second-order stochastic dominance, Assumption~\ref{as3} is in fact only satisfied asymptotically, which seems to be not addressed extensively in some treatments of stochastic dominance like e.g., \cite{Abadie}, (cf., also \cite{Barrett}, particularly footnote 11 at page 78)  and may be a point of interest for itself. Finally, note that the literature on stochastic dominance (e.g., \cite{Abadie,Barrett}) usually preferes to use a bootstrap test instead of a permutation test.
\subsection{Testing with Contaminated Protocols}\label{cont_prot}
\noindent The assumptions we made in the previous section in order to construct an (asymptotically) valid statistical test can be roughly divided into two categories: Assumptions~\ref{as1}, \ref{as2}, and \ref{as3} concerned the nature of the choice function under consideration, whereas Assumption~\ref{as4} concerned the process the protocol which the test is based upon was collected: \textit{i.i.d.}~for each fixed action description. Consequently, the validity of Assumption~\ref{as4} is often out of the control of the person analyzing the EDP and therefore questionable in practical situations. In order to protect against deviations from an idealized i.i.d.~scenario, we want to modify our statistical test from Theorem~\ref{theo2} so that it remains (asymptotically) statistically valid even under (rather small) deviations from this idealization. In other words, we aim for a modification of our test that still respects the given level $\alpha$ even under (slight) violations of Assumption~\ref{as4}. We achieve this by weakening the assumption: The idea here is no longer to assume that the consequences following a specific action description are generated i.i.d., but instead to allow them to follow arbitrary patterns of distribution and dependency. The concrete relaxation then arises indirectly via a restriction to the empirical distributions of the consequences following an action description: We assume that under the null hypothesis of our modified test, the empirical distribution under i.i.d. lies within a (quantifiable) neighborhood of the actually observed (potentially contaminated) empirical distribution. Formally, we arrive at:
\begin{as}\label{as5}
    For every $X_{a_i} \in \mathcal{D}$, the sub-protocol $\mathscr{P}_{\mathbf{z}}(\{i\})$ is assumed to be a contaminated version of an i.i.d.~sample $(c_k)_{k \in \underline{z_i}}$ from the distribution $\pi_{X_{a_i}}$. Precisely, for every contaminated version $\tilde{\mathbf{c}}:=(\tilde{c}_s)_{s \in \underline{q}}$ of a sample $\mathbf{c}:=(c_s)_{s \in \underline{q}}$ from a distribution $\xi$ on $C$, we can specify a credal set $\mathcal{
    M}(\tilde{\mathbf{c}},\xi)$ of probability distributions that contains both the empirical distributions based on $\tilde{\mathbf{c}}$ and $\mathbf{c}$.
\end{as}
The idea for a modified version of our test from Theorem~\ref{theo2} that remains valid when Assumption~\ref{as4} is replaced by Assumption~\ref{as5} is immediate: In each step, we compare the \textit{most conservative} value of the statistic that is compatible with the respective credal sets based on the observed sample to the \textit{most optimistic} value of the statistic that is compatible with the respective credal sets of the corresponding synthetic sample from the resampling process. 
  The exact method used to calculate the most conservative or optimistic value of $T$ naturally depends on the test statistic used (and the nature of its dependence on the various empirical distributions). Directly building on Theorem~\ref{theo2}, we now illustrate this robustification for EU-based choice functions as in Definition~\ref{EUbased}.
%
\begin{theorem} \label{theo3}
Let $(\mathbb{A},\mathscr{P}_{\mathbf{z}})$ and $\mathcal{F}$ be defined as in Theorem~\ref{theo2} and let $ch:=ch_{\mathcal{F}}$ be as in Definition~\ref{EUbased}. Moreover, let Assumptions~\ref{as1}, \ref{as2}, \ref{as3}, and \ref{as5} hold true. For suitable probability measures $\pi' , \pi''$, and suitable $C$-valued variables $U,V,$ set: 
$$D_{U,V}(\pi' , \pi''):= \inf\nolimits_{f \in \mathcal{F}} \Bigl\{\mathbb{E}_{\pi'}(f \circ U) -\mathbb{E}_{\pi''}(f \circ V)\Bigr\}$$
Using Assumption~\ref{as5}, define the value
$$d^{ji}_0:= \sup\Bigl\{D_{X_{a_j},X_{a_i}}(\pi' , \pi''):\pi'\in \mathcal{
    M}(\mathscr{P}_{\mathbf{z}}(\{j\}),\pi_{X_{a_j}}) , \pi'' \in \mathcal{
    M}(\mathscr{P}_{\mathbf{z}}(\{i\}),\pi_{X_{a_i}})\Bigr\},$$
and denote by $Q(\alpha,j,i)$ the (empirical) $\alpha$-quantile of the values
    $$\hspace{-0.6cm}\Biggl(\inf\Bigl\{D_{X_{a_j},X_{a_i}}(\pi' , \pi''):\pi'\in \mathcal{
    M}\bigl((w_{\sigma(s)})_{s \in \underline{z_j}},\pi_{X_{a_j}}\bigl) , \pi'' \in \mathcal{
    M}\bigl((w_{\sigma(s)})_{s=z_j+1,\dots , \ell},\pi_{X_{a_i}}\bigl)\Bigr\}\Biggr)_{\sigma \in \Pi(\underline{\ell})},$$
    where $\Pi(\underline{\ell})$ is the set of all permutations of the set $\underline{\ell}$ and $\ell:= z_i+z_j$ and $(w_1 , \dots , w_{\ell}):=(\mathscr{P}_{\mathbf{z}}(\{j\}),\mathscr{P}_{\mathbf{z}}(\{i\}))$.~Then, rejecting $H_0$ from Equation~\eqref{global} whenever $d^{ji}_0< Q(\alpha,j,i)$ for all $j$ with $X_{a_j} \in \mathcal{D}\setminus \{X_{a_i}\}$ gives an asymptotically valid level $ \alpha$ test for $(H_0,H_1)$.
\end{theorem}
\noindent \textit{Proof.} 
See Appendix.
\hfill $\square$

%
%
\section{Direct Credal Inference via Noise-robust ECFs Using Contamination Models}\label{direct_inference}
\noindent In Section~\ref{estimation}, we saw that the empirical variant of the EU-based ECF acts as a consistent approximator of the population variant of the choice function under certain conditions. While this is certainly reassuring from a purely mathematical point of view---at least in the limit, one is doing the right thing---it still provides little information about the uncertainty associated with using the empirical choice function in the finite sample case. This is why we formalized the dependence of the ECF on the intended contamination of the underlying empirical distributions in Section~\ref{testing}. 
We then turned to statistical permutation tests to test the significance of the choice sets obtained by the EU-induced ECF using appropriate statistical tests. Credal sets proved useful in this endeavor, as they allowed modeling contaminates samples, see Assumption~\ref{as1} and Theorem~\ref{theo3}. In what follows, we aim to robustify empirical choice function in a more direct way, but still relying on credal sets.    
\\[.2cm]
Empirical decision problems do not hinge on strong population-level assumptions like SWA and CWA. They allow for what we refer to as \textit{direct inference}: We do not, as is customary, relate the sample to the population via common \textit{i.i.d.}-style assumptions and then draw conclusions in an indirect way. Instead, we \textit{directly} approximate the underlying population choice function by the empirical choice function, see Definitions~\ref{def:ecf} and~\ref{def:consistent-approx}.
\\[.2cm]
This comes at the price of a stronger dependence on the \textit{observed} protocols. Through an epistemiological lense, \textit{direct inference} might be susceptible to what Horkheimer and Adorno \citep[page~vii]{horkheimer_adorno_dda_engl} refer in general to as the positivist "myth of that which is the case": adhering to observations without questioning their origin and relation to the \textit{explanandum}, see also \cite{herrmann2024position,pos-25} in an explicit statistical and machine-learning context. Surely, this is not desirable. In order to hedge against these risks of blind positivism, we aim at a nuanced robustification of ECFs. In particular, we want to robustify emiprical variants of the ECFs towards noise in the observed protocols. 
%
%
\\[.2cm]
\noindent In what follows, we specifically build on \textit{$\gamma$-contamination models} (or \textit{linear-vacuous models}, see, e.g.,~\cite[p.~147]{Walley:1991} and our Introduction).
Recall we implicitly assumed that there exists some probabilit  $\pi$ producing the latent states of the world $s \in S$, making $\mathscr{P}_{\mathbf{z}}(\{i\})$ a sample from $\pi_{X_{a_i}}$. Denote its empirical variant by $\hat \pi_{X_{a_i}}$. The corresponding \textit{$\gamma$-contamination model} is specified via the credal set
\begin{equation} \label{lvm}
\mathcal{M}_{X_{a_i}}(\gamma) =\Bigl\{ \xi_{X_{a_i}}: \xi_{X_{a_i}} \geq (1 -\gamma)\cdot \hat \pi_{X_{a_i}}\Bigr\},
\end{equation}
where $\xi_{X_{a_i}}$ is a probability measure, $\geq$ is understood event-wise and $\gamma \in [0,1]$. For examples of how to compute test statistics based on $\gamma$-contamination models, refer to \citet{uaiall}. Note that this specific $\gamma$-contamination model shall primarily serve as an illustration, see Section~\ref{testing} for more general contamination models. Further recall that we defined an empirical choice function $\hat{\text{ch}}_{\mathbf{z}}: S(\mathscr{P}_{\mathbf{z}})\to 2^{\mathbb{A}}$ via the requirement $\hat{\text{ch}}_{\mathbf{z}}(\mathscr{P}_{\mathbf{z}}(M)) \subseteq \{a_i: i \in M\}$ for all $M \subseteq \underline{n}$. A very natural way to directly robustify an empirical choice function is to require its output to be a superset of $\hat{\text{ch}}_{\mathbf{z}}(\mathscr{P}_{\mathbf{z}}(M))$, as detailed now.    

\begin{defi}\label{def:recf}
    Let $(\mathbb{A},\mathscr{P}_{\mathbf{z}})$ be an EDP and $\hat{ch}_{\mathbf{z}}$ an ECF on $S(\mathscr{P}_{\mathbf{z}})$. A function 
$$\widehat{\text{rch}}_{\mathbf{z}}: S(\mathscr{P}_{\mathbf{z}})\to 2^{\mathbb{A}}$$ 
satisfying $\hat{\text{ch}}_{\mathbf{z}}(\mathscr{P}_{\mathbf{z}}(M)) \subseteq \widehat{\text{rch}}_{\mathbf{z}}(\mathscr{P}_{\mathbf{z}}(M)) \subseteq \{a_i: i \in M\}$ for all $M \subseteq \underline{n}$  shall be called \textbf{robustified empirical choice function (RECF)}. It is the robustified version of $\hat{ch}_{\mathbf{z}}$.
\end{defi}
\noindent In order to come up with a noise-robust ECF using contamination models, we first need to introduce a specific type of ECFs, namely such ECFs whose subprotocols can be characterized as a sample from $\pi$ producing the latent states of the world $s \in S$.

\begin{defi}\label{def:distr-ecf}
    Let $(\mathbb{A},\mathscr{P}_{\mathbf{z}})$ be an empirical decision problem and $\Pi_{\mathbb A}$ be a set of probability measures denoted by $\xi_{X_{a_i}}$. A \textbf{distribution-dependent empirical choice function} is a function
    \[
    \widehat{\text{dch}}_{\mathbf{z}}: S(\mathscr{P}_{\mathbf{z}}) \times \Pi_{\mathbb A} \to 2^{\mathbb{A}}
    \]    
    satisfying $\widehat{\text{dch}}_{\mathbf{z}}(\mathscr{P}_{\mathbf{z}}(M), \xi_{X_{a_i}}) \subseteq \{a_i: i \in M\}$ for all $M \subseteq \underline{n}$ and $\xi_{X_{a_i}} \in \Pi_{\mathbb A}$ for all $i \in M$. 
\end{defi}
\noindent Finally, we can define a robust distribution-dependent empirical choice function. Note that it can, but generally does not need to hold $\mathscr{P}_{\mathbf{z}}(\{i\})\sim  \hat \pi_{X_{a_i}}$ for all $i \in M$. In other words, we can apply $\widehat{\text{dch}}_{\mathbf{z}}$ also to other, potentially disturbed probability measures. This generality of Definition~\ref{def:distr-ecf} allows for a definition of robust distribution-dependent empirical choice functions.

\begin{defi}\label{def:robust-distr-ecf}
  Let $(\mathbb{A},\mathscr{P}_{\mathbf{z}})$ be an EDP and $\widehat{dch}_{\mathbf{z}}$ a distributional empirical choice function on $S(\mathscr{P}_{\mathbf{z}})$.
  We call this distributional empirical choice function $\gamma$-\textbf{robust} 
  if and only if
\[\widehat{\text{dch}}_{\mathbf{z}}(\mathscr{P}_{\mathbf{z}}(M),  \hat \pi_{X_{a_i}} ) \subseteq \widehat{\text{dch}}_{\mathbf{z}}({\mathscr{P}}_{\mathbf{z}}(M),\xi_{X_{a_i}}) \subseteq  \{a_i: i \in M\} \] 
for all $M \subseteq \underline{n}$ and ${\mathscr{P}}_{\mathbf{z}}(\{i\}) \sim \xi_{X_{a_i}} $ for all $i \in M$ and for all $ \xi_{X_{a_i}} \in \mathcal{M}_{X_{a_i}} (\gamma)$.
\end{defi}
%
In words, Definition~\ref{def:robust-distr-ecf} views robustness as an outer expansion of the empirical choice set: for any $M$, the robust set keeps all action descriptions already selected by $\hat{\mathrm{ch}}{\mathbf z}$ and may add further non-rejectable ones.
Self-evidently, a $\gamma$-\textbf{robust empirical choice function} satisfies Definition~\ref{def:recf} of a robustified empirical choice function. On a more general level, it becomes evident that $\mathcal{M}_{X_{a_i}}$ from Equation~\eqref{lvm} with appropriate $\gamma$ constitutes a special case of the credal sets constructed according to Assumption~\ref{as5}. We can transfer Theorem~\ref{theo3} to the direct inference scenarios as follows.

\begin{cor}\label{cor:direct}
    Consider again a choice function $ch = ch_{\mathcal{F}}$ induced by an expectation orders, see Definition~\ref{EUbased}. The test described in Theorem~\ref{theo3} is an asymptotically valid level $\alpha$ test for $(H_0, H_1)$ from \eqref{global} under Assumption~\ref{as1}, if the credal sets $\mathcal{
    M}(\mathscr{P}_{\mathbf{z}}(\{i\}),\pi_{X_{a_i}})$ are replaced by $\mathcal{M}_{X_{a_i}}$ for corresponding $\hat \pi_{X_{a_i}}$ with $\gamma$ large enough such that the empirical measure of the
    i.i.d.~sample is contained.  
\end{cor}
\noindent In other words, the validity of our statistical permutation test is not affected by whether we center the credal set around the contaminated or the \textit{i.i.d.} sample, as long as the credal set contains both. Notably, the setup of $\gamma$-robust distribution-dependent empirical choice function paves the way for direct credal inference beyond the special cases of choice functions induced by expectation orders. We leave this to future work.




%
\section{Proof of Concept: EDPs for Evaluating Prompting Strategies for ChatGPT}\label{application}
%
%
\noindent We now come to an (illustrative) application of our theory in the context of comparing prompting strategies when using ChatGPT. Specifically, we want to use our framework to study whether the degree of politeness in prompts has an impact on the quality of the responses produced. Here, we operationalize quality by a two-dimensional metric composed of \textit{perplexity (PPL)} \citep{10.1121/1.2016299,holtzman2019curious} that measures how well a probabilistic model predicts a sequence of words, and \textit{coherence} \citep{10.1121/1.2016299} that calculates the log-likelihood of generated text given its prompt (thus, low perplexity values are desirable while for coherence it's vice versa).\footnote{For a comparison of quality metrics for text generation, see, e.g., 
\cite{garces-arias-etal-2025-towards,arias2025decoding,arias2025statisticalmulticriteriaevaluationllmgenerated}} It should be noted that although prompt politeness has recently received quite some attention in connection with the energy consumption of LLMs \citep{delavande2025thankyouenergy,raebsamen2025politenessai}, this section serves more as a proof of concept than a fully developed application. However, we believe that our example can serve as a blueprint for future studies in this direction.
\subsection{Setup of the Study}
\noindent We want to exemplarily compare three different degrees of politeness in prompting. These correspond exactly to the different action descriptions in the underlying EDP. The common context of the action descriptions in our EDP is the search for the best possible explanation of the board game GO with a fixed length. Concretely, we have:
\begin{itemize}
    \item[] $a_1$: \textbf{neutral} (\textit{``Explain GO in exactly 20 words.''})
    \item[] $a_2$: \textbf{polite} (\textit{``Explain GO in exactly 20 words, please.''})
    \item[] $a_3$: \textbf{inpolite} (\textit{``Tell me what GO is in exactly 20 words. No more, no less. Just do it.''})\end{itemize}
 Our protocol is structured as follows: Each of the three action descriptions was applied to the free version OpenAI’s ChatGPT model GPT-5 multiple times ($a_1$ eleven times, $a_2$ fourteen times, and $a_3$ ten times).\footnote{The experiment was performed on August 10, 2025.} As a consequence of each execution, the vector (PPL,Coh) was calculated from perplexity and coherence. The corresponding protocol is shown in Tables \ref{prot_neutral} to \ref{prot_polite} in the Appendix. As the choice function, we select the ECF based on the FSD rule $ch_{\succsim , \pi}$ in 
 Equation~\eqref{FSD}. The set $\mathcal{F}$ is then the set of all uniformly bounded functions that are isotone with respect to the component-wise order on $\mathbb{R}^2$. Specifically, we want to answer the following three questions:
\begin{itemize}
    \item[] $\mathbf{Q_1}$: How does the choice set look like if we apply the empirical variant of \eqref{FSD} as an ECF? 
    \item[] $\mathbf{Q_2}$: Can we use our test to derive inferential guarantees for the choice sets of \eqref{FSD}?
    \item[] $\mathbf{Q_3}$: How stable are these guarantess under (small) deviations from the \textit{i.i.d.} assumption?
\end{itemize}
\subsection{Study and Results}
\begin{figure}
    \centering
\includegraphics[scale=.42]{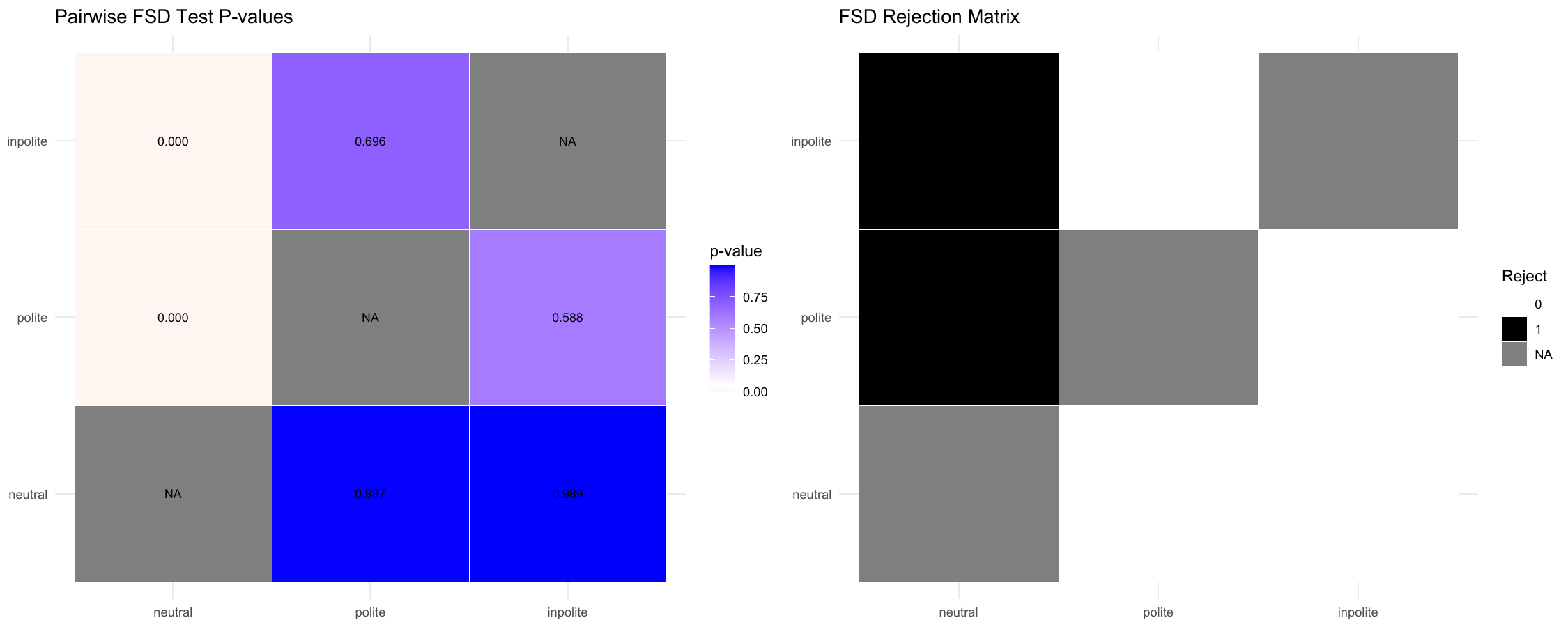}
    \caption{\textbf{Left:} The (approximated) p-values of all six pairwise tests on FSD. \textbf{Right:} The test decisions based on these p-values (where black encodes a reject decision).}
    \label{fig:pvals}
\end{figure}

\begin{figure}
    \centering
\includegraphics[scale=.52]{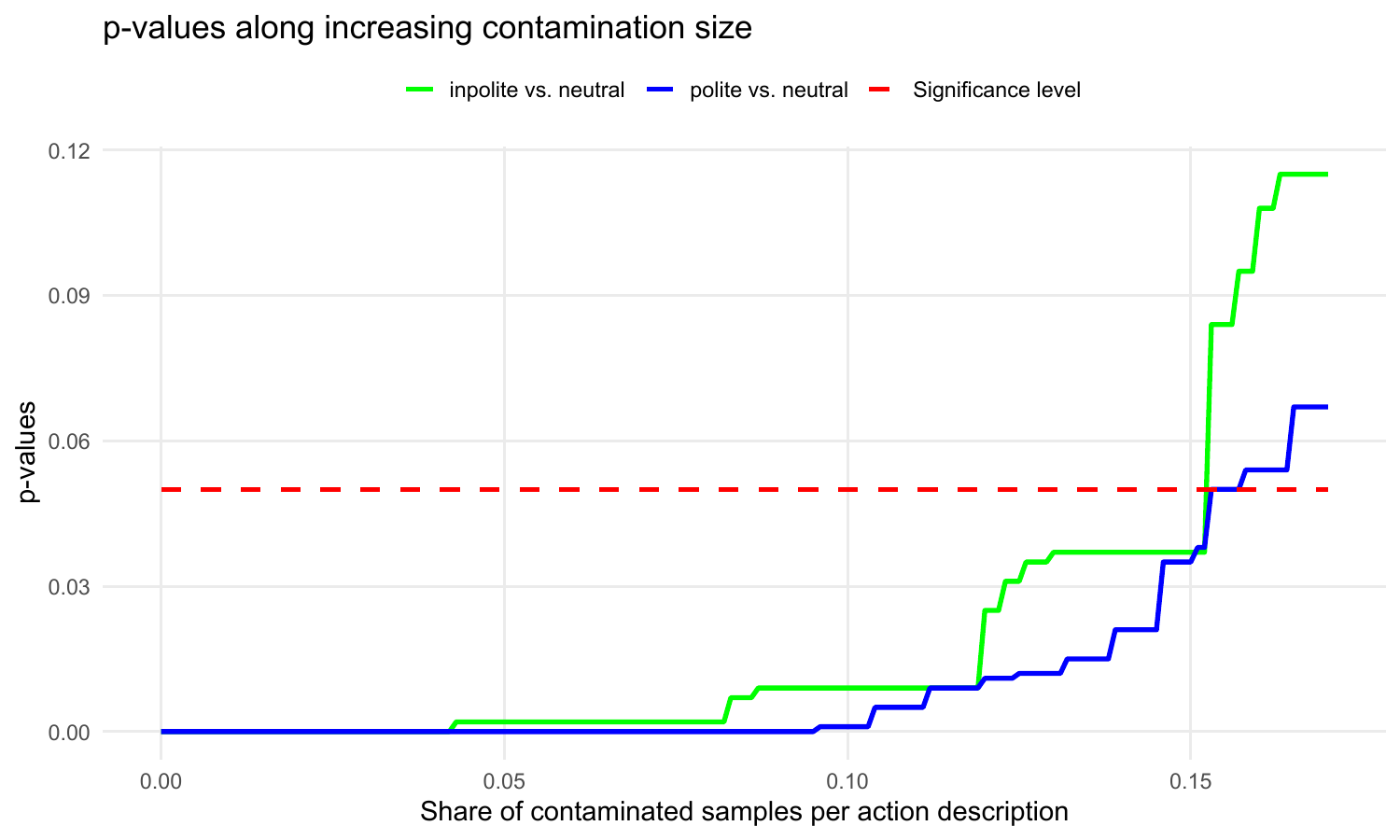}
    \caption{The p-values of the two pairwise FSD-tests along increasing share of contaminated data points (observed consequences) in the underlying protocol. The dashed red line represents the significance level $\alpha=0.05$.}
    \label{fig:cont}
\end{figure}

\noindent To answer $\mathbf{Q_1}$, we pairwisely calculate the test statistic $T_{U,V}$ from Theorem \ref{theo2} for all relevant subprotocols in our protocol. For example, we check whether the consequence sample given in Table \ref{prot_neutral} based on $a_1$ empirically FSD-dominates the consequence sample given in Table \ref{prot_polite} based on $a_2$. Specifically, we perform these calculations by evaluating the linear program proposed in \citet[Proposition 4]{uaiall}. It shows that all three prompting strategies are incomparable with respect to the empirical variant of FSD. Therefore the empirical choice set of this ECF does not exclude any of the three strategies, suggesting that politness in prompting does not have much effect on the quality of the generated text and, thus, neutral prompting is just fine (and presumably the best option in the light of energy consumption).
\\[.2cm]
To answer $\mathbf{Q_2}$, i.e., to move to inferential statements, we perform the hypothesis test described in Theorem \ref{theo2}~iii) to check whether we can significantly exclude being neutral in prompting from the choice set of population-based FSD as described in Equation~\eqref{FSD}. Concretely, we test
$$ H_0: X_{neutral} \notin ch_{\succsim , \pi}(\mathcal{D}) ~~ \text{ vs. }~~ H_1: X_{neutral} \in ch_{\succsim , \pi}(\mathcal{D})$$
with $\mathcal{D}$ being defined as $\{X_{neutral},X_{polite},X_{inpolite}\}$. Technically, for evaluating the test statistic for both the samples actually observed as well as their permuted counterparts, we again use the linear programs proposed in \citet[Proposition 4]{uaiall}. We perform the test for $N=10000$ randomly drawn permutations. A glance at Figure \ref{fig:pvals} immediately reveals that only two of the six pairwise tests (polite vs. neutral and inpolite vs. neutral) are significant at a level of $\alpha=0.05$. We can therefore reject the hypotheses that either polite prompting or impolite prompting significantly outperforms neutral prompting. Due to Theorem \ref{theo2}, it can now be concluded that the hypothesis $H_0: X_{neutral} \notin ch(\mathcal{D})$ formulated above must also be rejected. In other words, at a level of $\alpha=0.05$, there is no statistical evidence to suggest that neutral prompts lead to significantly worse results.
\\[.2cm]
To answer $\mathbf{Q_3}$, i.e., to investigate how stable our significant test decision from $\mathbf{Q_2}$ remains under small deviation from the \textit{i.i.d.} assumption, we want to apply the robustified statistical test described in Theorem \ref{theo3}. For this, we first need to specify the sample  contamination procedure described in Assumption \ref{as5}: we assume that for each action description $a_i$ the corresponding subprotocol is no longer i.i.d., but that the true underlying i.i.d.~sample does not deviate by more than $k$ data points from the actually observed sample. Formally, this means that for each sample of consequences $\tilde{\mathbf{c}}:=(\tilde{c}_s)_{s \in \underline{q}}$ we observe by our generating , we assume that the empirical distribution of the corresponding i.i.d.~sample lies in the contamination set $\mathcal{M}_{X_{a_i}}(\tfrac{k}{s})$ as defined in Equation~\eqref{lvm}.
\\[.2cm]
%
\noindent For this situation, it has been shown in \citet[A 3.3]{jansen2024statistical} that the p-value of the robustified test proposed in our Theorem \ref{theo3} can be written as a function in the contamination parameter $k$, given the evaluations of the test statistics for the actual observed samples as well as their permuted counterparts. In other words, for this specific contamination model, the robustified test can be obtained from the non-robust one without any significant increase in computational demand. We therefore apply the result from \citet{jansen2024statistical} for analzing dependence of our significant test decision as a function of $k$ (or, more generally, as  function of the \textit{share} of contaminated data points). The result is that the decision that neutral prompting significantly lies in the choice set of FSD remains significant even if \textit{about 15\% of the sample points per action description} from our protocol violates the \textit{i.i.d.} assumption in any arbitrary way.  Precisely, while the significance of the decision \textit{inpolite vs.~neutral} breaks from contamination sizes of slightly above 15\% on, the decision \textit{polite vs.~neutral} does so for contamination sizes slightly above 16\%. It can therefore be concluded that although the test decision in the present case exhibits a certain robustness with regard to the modeling assumptions, it is nevertheless relatively dependent on precisely these assumptions. An illustration of the test result is given in Figure~\ref{fig:cont}.





\subsection{Discussion}
\noindent Before we move on to discussing the results of our application, we should first again emphasize its illustrative nature: The results are not readily reproducible, as it is e.g.~not possible to set a seed when using the free version of ChatGPT. Furthermore, our linguistic operationalizations of politeness levels are ad hoc, and there is certainly room for discussion about the fluid boundaries between for instance neutral and polite language. We are by no means experts in this field and do not claim to be. 
In short, our application is a proof of concept that should not be interpreted in terms of content. That said, there are still some interesting findings that can be gleaned from our proof of concept: Firstly, this opens up a truly exciting potential area of application of our theory. Once the pitfalls mentioned above have been cleared out of the way and experts from the relevant fields have been consulted, it will be possible to look at the quality analysis of generative AI models from a completely new perspective. This is particularly exciting because, although reliable evaluation of such models is already important, it will certainly become even more essential in the future. Trustworthiness appears in two different facets in our theory: First, our theoretical results on estimating and testing choice functions (see Theorems~\ref{theorem_consistent_approximator} and~\ref{theo2}) allow us to backbone quality statements about generative AI models with inferential guarantees. This means that it is no longer just the quality of the generated protocol that is assessed, but rather the quality of the model as such (within the scope of the specified requirements, of course). The second facet refers to the possibility of examining the robustness of the guarantees obtained under deviations from the underlying statistical assumptions (see Theorem~\ref{theo3}). More precisely, we can explicitly examine how strongly our test decisions depend on the assumption 
of i.i.d.~protocols. Such a robustness analysis is extremely interesting in the context of our application: If the same prompt is used multiple times in succession for an LLM, a dependency structure will emerge between the responses received (and thus their quality), depending on the model used. It can therefore be assumed with a high degree of certainty that 
the i.i.d.~assumption on which the validity of our permutation test is based is not fulfilled. In our example, however, we see that our test decisions would still remain statistically valid if 15\% of the observed consequences per action description came from any distribution. This substantially secures the validity of the inferential analysis and also offers the possibility of questioning guarantees if they do not appear robust enough.
\section{Summary, Discussion, and Future Research}
\subsection{Summary}
\noindent In this article, we introduced empirical decision problems as an alternative to classical decision theory that is based purely on observational data (protocols) and thus avoids the strong structural assumptions of the classical theory such as CWA or SWA. We were able to show that the central decision-making tool of classical decision theory -- choice functions -- can also be adapted to the empirical case discussed here. We could see that this adaptation has the advantage that many concepts can be transferred more or less directly from classical theory to our setting. Examples include the transfer of common decision-making rules such as expected utility maximization or first-order stochastic dominance. However, basing our theory purely on observational data naturally has its price: any statement about optimal actions refers purely to their empirical optimality given the observations made so far. In other words, we initially make a purely descriptive statement about the observed protocol, and not about the optimality of the respective action description in the underlying population.
\\[.2cm]
In order to relate the optimality statements derived from an EDP to the underlying population, we use statistical learning theory on the one hand and non-parametric statistical methods on the other: In Theorem~\ref{theorem_consistent_approximator}, we show how, under mild regularity assumptions on the consequence space, ECFs can be constructed such that they form consistent approximators of underlying population choice functions. In Theorem~\ref{theo2}, we show how, given an EDP, a consistent statistical test can be constructed that examines whether an action belongs to a population choice set. Finally, in Theorem~\ref{theo3}, we demonstrate how the use of ideas from the theory of imprecise probabilities can be used to make our test more robust to deviations from the \textit{i.i.d.} assumption. In other words, our inferential guarantees go far beyond the usual, allowing us to analyze the dependence of their validity on the assumptions made about the design of the protocols.
\\[.2cm]
The decision-making framework presented here, based solely on observable components, has many potential areas of application. Specifically, we have focused on a proof-of-concept application in the field of generative AI model evaluation, the results of which look extremely promising: To the best of our knowledge, both the inferential guarantees of the evaluations resulting from our framework and their robustness analyses are rather novel in this currently much-discussed field (an exception is given by \cite{arias2025statisticalmulticriteriaevaluationllmgenerated}).
\subsection{Discussion and Future Research}
\noindent While we are strongly convinced of the usefulness of the new perspective on decision-making under uncertainty described here, we should not ignore the limitations of our framework and the avenues for future research that arise from them. The first major limitation has already been discussed and illustrated in Section~\ref{limitations}: The analysis of EDPs, or at least the inferential and robustness guarantees derived from them, only apply in frequentist settings. As soon as we doubt the randomness of the generation of states and suspect a deterministic, possibly even hostile mechanism behind them, our theory remains silent about these situations. In such cases, it would be much more appropriate to first try to learn the deterministic mechanism from a protocol and then use this knowledge to develop the best decision-making strategies. However interesting this may be, it would cover a completely different set of questions and thus go far beyond the scope of the results presented in this paper. Interestingly,  these problems could still be discussed—at least in purely formal terms—within the same framework proposed here. This opens up a interesting avenue for future research.
\\[.2cm]
Another limitation concerns the design of the protocols on which our theory is based. To ensure the validity of our inferential guarantees (Theorems~\ref{theorem_consistent_approximator} and~\ref{theo2}), we assume that the consequences under each individual action description can be interpreted as an i.i.d.~sample. For various reasons, this is questionable in practice, to say the least. On the one hand, in real-world problems, the state space is likely to be dynamic, so that, strictly speaking, each time an action description is followed, one could formally be faced with a slightly different decision problem. However, even if we assume a static state space, the assumption that dependencies do not arise when an action description is repeated multiple times seems rather strong in many situations. To a certain extent, our framework is able to respond adequately to these challenges: In Theorem~\ref{theo3}, we have shown that at least the inferential guarantees obtained through testing can be extended to protocol designs that deviate from a perfect i.i.d.~structure. However, it should be clearly noted here that our framework is more suited to handling isolated outliers or sampling processes that deviate only slightly from an i.i.d.~design. Our framework will only very rarely produce interesting statements about designs substantially differing from i.i.d., as the modified statistical test from Theorem~\ref{theo3} will rapidly lose much of its power in such cases. It is definitely a very interesting avenue of research to investigate whether stronger statements can be made here, for example by first attempting to learn the dependency structures from the protocols.

\section*{Acknowledgements}
\noindent We would like to thank the participants of the Workshop on Learning under Weakly Strucuted Information (LuWSI)~2025 in Tübingen and the International Symposium on Imprecise Probabilities: Theories and Applications (ISIPTA)~2025 in Bielefeld (where we presented some initial ideas as a poster, see~\cite{jansenempirical}) for many interesting discussions and suggestions. JR acknowledges support by the Bavarian Academy of Sciences and the LMU Mentoring Program. 
\section*{Appendix}
\subsection*{A1. Proofs of the Theorems}
\subsection*{Proof of Theorem~\ref{theorem_consistent_approximator}}
\noindent First, consider $cr(\cdot)_1$: For random variables $X=X_{a_i}$ , $Y=X_{a_j}$ and their empirical analogues $\hat{X},\hat{Y}$ we have $\mathfrak{d}(X,\hat{X})\stackrel{a.s.}{\longrightarrow} 0$ and $\mathfrak{d}(Y,\hat{Y})\stackrel{a.s.}{\longrightarrow} 0$ as 
$z_1$ or $z_2$  goes to infinity, respectively, because $\mathfrak{C}$ has finite VC dimension. (For this part, it would suffice to assume that $\mathfrak{C}$ is a Glivenko-Cantelli class w.r.t. the underlying probability measure.) Therefore, if $\mathbf{\underline{z}}$ goes to infinity, almost surely both pseudodistances go simultaneous to zero and since $cr$ is continuous, we have

\begin{align}
cr(X,Y)_1 > 0 &\Longrightarrow \mbox{ a.s.: }\exists z_0\in \mathbb{N}: \forall \mathbf{\underline{z}} \geq z_0 :cr_1(\hat{X},\hat{Y}) >0.\label{cr1}
\end{align}
\noindent Second, consider $cr(\cdot)_2$: With the same argumentation as above, it immediately follows
\begin{align}
cr(X,Y)_2 > 0 &\Longrightarrow \mbox{ a.s.: }\exists z_0\in \mathbb{N}: \forall \mathbf{\underline{z}} \geq z_0 :cr_2(\hat{X},\hat{Y}) + 4 \delta(\varepsilon_{\mathbf{\underline{z}}}) \geq 0
\end{align}
Now, consider the case $cr(X,Y)_2=0$: Since $\mathfrak{C}$ has finite VC dimension, it follows from \cite[p. 192 Theorem 5.1]{vapnik1998statistical}, that for arbitrary $\xi>0$: 
\begin{align*}
    P\left( \sup\limits_{f\in \mathcal{F}} \left|\mathbb{E} (f \circ X) - \hat{\mathbb{E}} (f \circ \hat{X}) \right| >\xi \right) \leq  4 \left(\frac{e\cdot 2z_i}{h}\right)^h \cdot \exp \left\{ -\xi_*^2 z_i\right\}\\
\end{align*}
where $\xi_*=\frac{\xi -1/z_i}{B-A}$; $A$ and $B$ is the upper and the lower uniform bound of $\mathcal{F}$; $h$ is the VC dimension of $\mathfrak{C}$ and $P$ denotes the infinite product law of the image law of $\pi$. (Note that for this statement it is not sufficient to only assume that $\mathfrak{C}$ is a Glivenko-Cantelli class w.r.t. the underlying probability measure.) Now, look at $X=X_{a_i}$ and set $\xi:=\varepsilon_{z_i}:= 1/\sqrt[4]{z_i}$. If $z_i$ is large enough, we have $\xi_* = \varepsilon(z_i) - 1/\mathbf{z_i}  \geq \varepsilon(z_i)/(2(B-A))$ and we get
\begin{align*}
      P(|cr_2(X,Y) - cr_2(\hat{X},Y)| > \delta(\varepsilon_{z_i}))   & \leq P\left( \sup\limits_{f\in \mathcal{F}} \left|\mathbb{E} (f \circ X) - \hat{\mathbb{E}} (f \circ \hat{X}) \right| >\varepsilon_{z_i} \right)\\
     &\leq  4 \left(\frac{e\cdot 2 z_i}{h}\right)^h \cdot \exp \left\{ -\xi_*^2 z_i\right\}&\\
     &\leq 8 \left(\frac{e\cdot 2z_i}{h}\right)^h \cdot \exp \left\{ -\varepsilon_{z_i}^2 z_i/(2(B-A))^2\right\}&\\
     &= 8 \left(\frac{e\cdot 2z_i}{h}\right)^h \cdot \exp \left\{ -\sqrt{z_i} /(2(B-A))^2\right\}&
     \end{align*}
Now, observe that the right-hand side of the last inequality converges very quickly to zero as $z_i$ goes to infinity, More concretely, the series $\sum\limits_{z_i=1}^\infty 8 \left(\frac{e\cdot 2z_i}{h}\right)^h \cdot \exp \left\{ -\sqrt{z_i} /(2(B-A))^2\right\} $ is finite. Thus, by the Borel-Cantelli lemma we have that almost surely there exists $z_0 \in \mathbb{N}$ such that $ |cr_2(X,Y) - cr_2(\hat{X},{Y})| \leq  \delta(\varepsilon_{z_i}) \leq\delta(\varepsilon_{\mathbf{\underline{z}}})$ for $z_i \geq z_0$. Note that above the convergence is analyzed in the sense of setting the index $l$ of the analyzed sequence to $l=z_i$. Therefore, a protocol with $l$ observed action descriptions $a_i$ occurs only once in this sequence. However, the actual sanpling process is such that at every step, a different action may be chosen such that there is usually more than one step, where $z_i=l$, but this does not change the convergence w.r.t. the whole process, because 'after' $z_0$, for different $z$ , $\tilde{z}$ with $z_i=\tilde{z}_i=l$ we still have $l\geq z_0$. Therefore, we also have that almost surely there exists $z_0 \in \mathbb{N}$ such that $ |cr_2(X,Y) - cr_2(\hat{X},{Y})| \leq \delta(\varepsilon_{\mathbf{\underline{z}}})$ for $\mathbf{\underline{z}} \geq z_0$. Now, an analogous statement also holds for $ |cr_2(X,Y) - cr_2(X,\hat{Y})| $ and therefore, with the help of the triangle inequality, we get that
%
%


\begin{align}
cr(X,Y)_2 > 0 &\Longrightarrow \mbox{ a.s.: }\exists z_0\in \mathbb{N}: \forall \mathbf{\underline{z}} \geq z_0 :cr_2(\hat{X},\hat{Y}) + 4 \delta(\varepsilon_{\mathbf{\underline{z}}}) \geq 0.
\end{align}
The same holds for the more general specification $\varepsilon_{\mathbf{\underline{z}}} \in \Theta\left( 1/\sqrt[4]{\mathbf{\underline{z}}}\right)$. All-together, this shows that if $X_{a_i}$ is not in a choice set $ch(\{X_{a_i},X_{a_j},\ldots\})$, then, almost surely $a_i$ is also asymptotically not in the empirical choice set $\hat{ch}_\mathbf{{z}}(\{a_i,a_j,\ldots\})$. It remains to show that if $X_{a_i}$ is in a choice set $ch(\{X_{a_i},X_{a_j},\ldots\})$, then, almost surely $a_i$ is also asymptotically in the empirical choice set $\hat{ch}_\mathbf{{z}}(\{a_i,a_j\ldots\})$. For this, look again at $X:=a_i$ and a further $Y:=a_j$ in the set $\mathcal{D}$. In the case $cr_1(X,Y) < 0$, we have, with the same argumentations like from above (cf., (\ref{cr1})) that almost surely $cr_1(\hat{X},\hat{Y}) <0 $ for $\mathbf{\underline{z}}$ large enough. In the case $cr_1(X,Y)=0$, because of the assumed antisymmetry of $cr$, we know that $cr_2(X,Y) <0$ and particularly $cr_2(X,Y) + \delta(\mathbf{\underline{z}}) <0 $ for large enough $\mathbf{\underline{z}})$ and therefore we analogously have that almost surely $cr_2(\hat{X},\hat{Y}) + 4 \delta(\mathbf{\underline{z}}) < 0 $ for $\mathbf{\underline{z}}$ large enough. This finishes the proof.
\hfill $\square$
\subsection*{Proof of Theorem~\ref{theo2}}
\noindent \textbf{i):} Statement $i)$ is true due to the construction of the permutation test: For a fixed pair comparison $X_{a_i}$ vs $X_{a_j}$, under i.i.d.~and $\tilde{H}_0$ the distribution of the resampled statistic (conditional on the observed values of the pooled sample) equals the (conditional distribution) of the test statistic under $\tilde{H_0}$. Therefore, the test for each pair comparison is a valid level $\alpha$ test and - due to its construction - the whole test is also a valid level $\alpha$ test.
\\[.3cm]
\noindent \textbf{ii)}  Since $\mathfrak{C}$ has finite VC dimension, under $\tilde{H}_0$, the distribution of the test statistics under the permutation scheme concentrate around $0$ and therefore the permutation based critical values (for fixed level $\alpha$) become arbitrary close to $0$ with arbitrary high probability as $\mathbf{\underline{z}}$ goes to infinity. On the other hand, for $X_{a_i} \nprecsim X_{a_j}$  we have $\inf\nolimits_{f \in \mathcal{F}}\mathbb{E}_\pi (f\circ X_{a_j}) - \mathbb{E}_\pi(f\circ X_{a_i}) = c_{ij} < 0$ and therefore, the corresponding test statistic almost surely converges to $c_{ij}$ which implies that for $X_{a_i} \in ch(\{X_{a_i},X_{a_j}\})$ almost surely there exists a $z_0$ such that ${a_i} \in \hat{ch}_{\mathbf{{z}}} (\{X_{a_i},X_{a_j}\})$ for $\mathbf{\underline{z}}\geq z_0$. Analogously, the same follows for protocols consisting of more than two action descriptions.
\\[.3cm]
\noindent \textbf{iii)} First, note that due to i) the permutation test has the property $\pi(T_N < c_N) =\pi (\bar{T}_N <c_N) \leq \alpha$ under the special case $\tilde{H}_0$, where $T_N$ is the test statistic for comparing $X_{a_i}$ with $X_{a_j}$, $\bar{T}$ is the permutet test statistic and $c_N$ is the critical value obtained with the permutation scheme. Now, if $c_N$ would be a constant, then one could argue that if Assumption~\ref{as3} is fulfilled exactly, then it would directly follow $\pi(T_N < c_N) \leq \pi( \bar{T}_N < c_N) \leq \alpha$ also under every arbitrary case of $H_0$. However, $c_N$ is not a constant, but a random variable (which is usually correlated with $T_N$), and this is important, especially for permutation tests, because the critical values of a permutation test are - by contsruction - tight in the sense that the event $T_N < c_N $ has probability bounded by $\alpha$ also conditional on the observed data values of the pooled sample. Unfortunately, the first-order stochastic dominance guaranteed by Assumption~\ref{as3} is generally not kept under conditioning (and assuming a conditional dominance instead of an unconditional dominance in Assumption~\ref{as3} seems to be a too hard to satisfy property), one cannot argue conditionally. However, one  can analyze conditions under which an appropriately normalized version $T_N$ of the test statistic, as well as the corresponding normalized permutet statistic $\bar{T}_N$, converge to a non-degenerate limit distribution $T^*$ and $\bar{T}^*$, respectively. 
If additionally the corresponding random (normalized) critical value $c_N$ of the permutation scheme converges (in probability) to a constant $c^*$ and if the distribution of both $T^*$ and $\bar{T}^*$ are continuous in a neighborhood of $c^*$, Assumption~\ref{as4}  guarantees asymptotic validity of the permutation test also under $H_0$, because then we have $$\alpha \geq \lim\limits_{N\to \infty}\pi (\bar{T}_N < c_N) =\pi(\bar{T}^* < c^*) \geq \pi(T^* < c^*) =\lim\limits_{N \to \infty} \pi(T_N < c_N).$$ %
Note that for such a proof strategy, only the asymptotic validity of Assumption~\ref{as3} is needed. Note further that there are cases where Assumption ~\ref{as3} is in fact only asymptotically valid, for example second-order stochastic dominance, compare Appendix~A3. This fact seems to be not very explicitly stated in other works such as, e.g., \cite{Abadie}(cf., also \cite{Barrett}) and may be a fact of its own interest. Now, let us establish the needed facts to establish the argumentation from above:\\[.15cm]
\noindent
Under the additional assumption of $iii)$, under ${H}_0$ - using the notation of \cite{van1996weak} -  the empirical process $\sqrt{m}(\bar{P}_m -H_N)$ converges weakly to $\sqrt{1-\lambda} G_H$ given almost every sequence $X_1,X_2, \ldots,Y_1,Y_2,\ldots$ cf., \cite[Theorem~3.7.2]{van1996weak}. Here $(X_1,\ldots,X_m)$ denotes the sample that corresponds to action dscription $a_i$ and $(Y_1, \ldots , Y_n)$ denotes the sample that corresponds to action description $a_j$. Furthermore $H_N$ is the empirical measure that corresponds to the pooled sample $(Z_1,\ldots, Z_N)=(X_1,\ldots ,X_m,Y_1,\ldots, Y_n)$ and $\bar{P}_m$ is the empirical measure of a random sample of size $m$ without replacement from the pooled sample. The proportion $\lambda$ is the limit $\lim\limits_{N\to \infty} z_i/(z_i+z_j)$ from Assumption~\ref{as4}. The limiting process $G_H$ is a (tight) $H$-Brownian bridge w.r.t. law $H = \lambda P + (1-\lambda) Q$ where $P=\pi_{X_{a_i}}$ and $Q=\pi_{X_{a_j}}$ (which by abuse of notation we identify with its associated Gaussian Radon measure on $\ell ^\infty(\mathcal{F}$)). Therefore, the  normalized permutet test statistic $\bar{D}_{m,n}:=\sqrt{mn/(N)}\inf\nolimits_{f\in \mathcal{F}} \left(\bar{P}_{m,N} - \bar{Q}_{n,N}\right)=1 /\sqrt{1-\lambda_N}\sqrt{m} \inf\nolimits_{f\in \mathcal{F}} \left(\bar{P}_{m,N} - \bar{Q}_{n,N}\right)$ (where $\lambda_N =z_i/(z_i+z_j)$ and $Q$ is the empirical measure associated to the remaining $n$ samples of the pooled sample) converges weakly to $\inf\nolimits_{f\in \mathcal{F}} G_H(f)$ (cf., \cite[p. 363]{van1996weak}). Now, the functional $S:\ell^\infty (\mathcal{F}) \to \mathbb{R}: z \mapsto -\inf\limits_{f \in \mathcal{F}} z(f)$ is convex. Thus, Theoorem~11.1 of \cite{davydov1998local} shows that $S\circ G_H$ has a continuous distribution function $F$ except for the separation point $r_0:=\inf\{r \mid F(r) >0\}$. (Note that $G_H$ is a Gaussian Radon measure, therefore the assumption of the theorem are actually satisfied.). Because $0\geq T=-S\circ G_H$  we have $S\circ G_H\geq 0$ and therefore $r_0\geq 0$. Actually, we have $r_0=0$, because the topological support of the Radon Gaussian measure $G_H$ is a linear subspace of $\ell ^\infty(\mathcal{F})$ (cf., \cite[p.~27]{bogachev2010differentiable})  and  therefore contains the null-function $f_0$ which implies $\pi(S\circ G_H \leq  \varepsilon )\geq \pi (||G_H -f_0 ||_\infty \leq \varepsilon) >0$ for arbitrary $\varepsilon>0$. 
Furthermore, we have $1\geq \pi(\forall f\in \mathcal{F}: G_H(f)\geq 0 \mbox{ or }\forall f\in \mathcal{F}: G_H(f) \leq 0)= \pi(\forall f \in \mathcal{F}: G_H(f)\geq 0) + \pi(\forall f \in \mathcal{F}: G_H(f)\leq 0) -\pi(\forall f \in \mathcal{F}: G_H(f)= 0)$. Now, because of Assumption~\ref{as4} the process $G_H$ is non-degenerate, and therefore the last summand is zero. Due to symmetry of $G_H$, the first two summands are both identical to $\pi(S\circ G_H)$, which gives $2 \cdot \pi(S\circ G_H=0)\leq 1$, which means that the seperation ppoint has maximal probability $1/2$. Since $T=-S\circ G_H$, for a significance level $\alpha < 1/2$, the point $r_0=0$ of possible non-continuity does not play a role asymptotically, since the normalized quantiles $c_N$ converge to $c^* < r_0=0$ for $\alpha >1/2$, It remains to show, that also the original statistic $T_N$ has a limiting distribution that is also continuous at least in a neighbourhood of $c^*$: Under $H_0$ we have $T_N=\sqrt{mn/N}\inf\limits_{f \in \mathcal{F}}(P_m f -Q_nf)$, where $P_m$ and $Q_n$ are the empirical measures for $X_{a_j}$ and $X_{a_i}$, respectively. Now, the process  $\sqrt{mn/N}(P_m -P -(Q_m-Q)) =
\sqrt{n/N}\sqrt{m} (P_m-P) +\sqrt{m/N} \sqrt{n}(Q_n-Q)$ converges weakly to an indpendent sum $\sqrt{1-\lambda}G_P - \sqrt{\lambda}G_Q$. 
Here, $P$ and $Q$ are the true image measures of $X_{a_j}$ and $X_{a_i}$, and $G_P$ and $G_Q$ are a tight $P-$Brownian bridge and a tight $Q$-Brownian bridge, respectively. Therefore the limiting process of $\sqrt{mn/N} (P_m-Q_m)$ is  the Gaussian Radon measure $\sqrt{1-\lambda}(G_P+P) - \sqrt{\lambda}(G_Q +Q)$. From this, analogous to above, it follows that the limiting distribution of $T_N$ is continuous except for possibly its separation point $r_1$. Further, we have $T_N=\sqrt{mn/N}\inf\limits_{f \in \mathcal{F}}(P_m f -Q_nf) =  \sqrt{mn/N}\inf\limits_{f \in \mathcal{F}} (P_mf -Pf-Q_fn+Qf +Pf -Qf) \geq \sqrt{mn/N}\inf\limits_{f \in \mathcal{F}}  ((P_mf - Pf) - (Q_mf-Qf)) + \sqrt{mn/N}\inf\limits_{f \in \mathcal{F}} (Pf-Qf) \geq \sqrt{mn/N}\inf\limits_{f \in \mathcal{F}} ((Pf_m - Pf) - (Q_nf-Qf))$. The last inequality is valid because under $H_0$ we have $\inf\limits_{f \in \mathcal{F}}(Pf -Qf) \geq 0 $. Now, the term 
$\sqrt{mn/N}((P_m -P)-(Q_m -Q))  = (\sqrt{n/N}\sqrt{m}(P_m-P) - \sqrt{m/N}  (Q_n-Q))$ converges weakly to the independent sum $(1-\lambda)G_P + \sqrt{\lambda}G_Q$. This independent sum is equal in distribution to $G_H$. 
Therefore, the limiting distribution of $T_N$ is stochastically larger than the limiting distribution of the permuted test statistic (which was asymptotically distributed as $ \inf\nolimits_{f\in \mathcal{F}} G_H(f)$). Therefore, the separation point $r_1$ of $T_N$ is larger than or equal to $r_0=0$. Therefore, also $T^*$ is in fact continuous in a neighborhood around $c^*$.  Importantly, note that the asymptotic stochastic dominance from above entails also that under the made Donsker class assumption, Assumption \ref{as3} also directly follows asymptotically. However, maybe the weaker assumption that $c^N$ converges to a value $c^*$ at which the distribution of $T_N$ and $\bar{T}$ have continuous distribution, plus Assumption~\ref{as3} can be proofed easier in certain situations.
\\[.2cm]
\noindent \textbf{iv)} This directly follows from Theorem~3.7.2 and Theorem~3.7.6 and the discussion after Theorem~3.7.7 (p.365) in \cite{van1996weak}.\hfill $\square$
\subsection*{Proof of Theorem~\ref{theo3}}
\noindent Under Assumption~\ref{as5}, the empirical measures of the (unknown) i.i.d.~samples are contained in the credal sets of the (known) contaminated samples. Since this holds for arbitrary samples, it applies specifically to every permutation sample. In a permutation test, the statistic is recomputed for each permutation of the combined sample and compared to its value on the original sample; larger resampled values provide evidence against the null. By Theorem~\ref{theo2}, this procedure is (asymptotically) valid under i.i.d.~sampling. In our setting, computing the statistic over all compatible empirical measures ensures that the resulting set contains the i.i.d.~statistic, both for the observed and for each permuted sample. To guarantee rejection only when fully justified, we adopt a worst-case rule: evidence against the null is counted only if the supremum of the observed statistic remains below the infimum of the corresponding resampled statistic. This coincides with the test from the theorem and therefore concludes the proof.
\hfill $\square$
\subsection*{A2. Results of the prompting experiments in Section~\ref{application}}
%
%
\begin{table}[h]
\centering
\scriptsize
\setlength{\tabcolsep}{4pt} 
\renewcommand{\arraystretch}{1} 
\begin{tabular}{r p{9.5cm} c}
\toprule
Trial & ``Explain GO in exactly 20 words.'' (\textbf{neutral}) & (PPL, Coh) \\
\midrule
1  & Go is a strategic board game from ancient China where two players aim to control territory using black and white stones. & (18, 0.95) \\
2  & Go is an ancient board game where two players place stones to control territory, requiring deep strategy, balance, and foresight. & (20, 0.92) \\
3  & Go is a two-player board game where black and white stones battle for territory using strategy, balance, and tactical positioning. & (19, 0.94) \\
4  & Go is a strategic board game where two players place stones to control territory, emphasizing balance, patience, and tactical depth. & (21, 0.91) \\
5  & Go is an ancient strategy board game where two players place stones to control territory through balance, patience, and tactical precision. & (22, 0.90) \\
6  & Go is an ancient board game where two players place stones on a grid, aiming to control the most territory with strategy. & (20, 0.93) \\
7  & Go is a strategic board game where two players take turns placing stones, aiming to control more territory than their opponent. & (19, 0.92) \\
8  & Go is an ancient board game where two players place black or white stones to control territory through strategy, patience, and positioning skill. & (23, 0.89) \\
9  & Go is an ancient board game where two players take turns placing stones, aiming to surround territory and outmaneuver their opponent strategically. & (21, 0.91) \\
10 & Go is a strategic board game where two players place stones on a grid, aiming to control the most territory by surrounding. & (18, 0.94) \\
11 & Go is a two-player board game where players place black or white stones to control territory by surrounding opponent’s stones strategically. & (17, 0.96) \\
\bottomrule
\end{tabular}
\caption{Responses to the prompt ``Explain GO in exactly 20 words.'' with estimated Perplexity and Coherence.}
\label{prot_neutral}
\end{table}
\FloatBarrier
\begin{table}[H]
\centering
\scriptsize
\setlength{\tabcolsep}{4pt}
\renewcommand{\arraystretch}{1}
\begin{tabular}{r p{9.5cm} c}
\toprule
Trial & ``Tell me what GO is in exactly 20 words. No more, no less. Just do it.'' (\textbf{inpolite}) & (PPL, Coh) \\
\midrule
1 & Go is a strategic two-player board game where black and white stones are placed on a grid to capture territory and outmaneuver opponents. & (19, 0.93) \\
2 & Go is a two-player board game where black and white stones are placed on a grid to control territory through strategic positioning. & (18, 0.94) \\
3 & Go is a strategic two-player board game where players take turns placing black or white stones to control territory and capture opponents’ stones. & (20, 0.92) \\
4 & Go is an ancient game where players place black and white stones on a grid, capturing territory through careful, strategic moves and positioning. & (21, 0.91) \\
5 & Go is a board game where two players place black and white stones on a grid, aiming to surround and capture territory. & (17, 0.95) \\
6 & Go is a complex board game where two players place black and white stones to claim territory, requiring strategy, patience, and foresight. & (22, 0.90) \\
7 & Go is an ancient two-player game where black and white stones are placed to control territory by surrounding the opponent’s stones strategically. & (20, 0.92) \\
8 & Go is a strategic board game where black and white stones are placed on a grid to capture territory and outmaneuver opponents. & (19, 0.93) \\
9 & Go is an ancient board game for two players who place black and white stones to control territory through strategy, balance, and patience. & (18, 0.94) \\
10 & Go is a strategic two-player board game where black and white stones are placed on a grid to control and surround territory. & (18, 0.93) \\
\bottomrule
\end{tabular}
\caption{Responses to ``Tell me what GO is in exactly 20 words. No more, no less. Just do it.'' with PPL and Coh.}
\label{prot_inpolite}
\end{table}
\begin{table}[h]
\centering
\scriptsize
\setlength{\tabcolsep}{4pt} 
\renewcommand{\arraystretch}{1} 
\begin{tabular}{r p{9.5cm} c}
\toprule
Trial & ``Explain GO in exactly 20 words, please.'' (\textbf{polite}) & (PPL, Coh) \\
\midrule
1  & Go is an ancient board game where two players strategically place black and white stones to control territory and outmaneuver opponents. & (18, 0.95) \\
2  & Go is an ancient strategic board game where two players place black and white stones to control territory and win. & (20, 0.92) \\
3  & Go is a two-player ancient board game where black and white stones are placed to control territory through strategic moves. & (19, 0.94) \\
4  & Go is a classic two-player board game where black and white stones are placed strategically to surround territory and capture opponents. & (21, 0.91) \\
5  & Go is an ancient two-player board game where players place black and white stones to surround territory and capture opponents’ stones. & (22, 0.90) \\
6  & Go is a strategic ancient board game where two players alternately place black and white stones to control territory and capture stones. & (20, 0.93) \\
7  & Go is an ancient board game where two players place black and white stones to surround territory and capture opponent stones strategically. & (23, 0.89) \\
8  & Go is a two-player strategic board game where black and white stones are placed on a grid to control territory and capture. & (19, 0.92) \\
9  & Go is an ancient board game where two players place black and white stones to control territory by surrounding and capturing stones. & (21, 0.91) \\
10 & Go is a two-player ancient board game where players place black and white stones to strategically surround territory and capture opponent stones. & (18, 0.94) \\
11 & Go is a strategic ancient board game where two players place black and white stones aiming to control territory by surrounding opponents. & (17, 0.96) \\
12 & Go is a strategic two-player board game where black and white stones are placed to surround and control territory on a grid. & (18, 0.95) \\
13 & Go is an ancient two-player board game where players place black and white stones to surround territory and capture opponents’ stones. & (20, 0.93) \\
14 & Go is a two-player strategic board game where black and white stones are placed to surround and control territory on a grid. & (19, 0.94) \\
\bottomrule
\end{tabular}
\caption{Responses to ``Explain GO in exactly 20 words, please.'' with estimated Perplexity and Coherence scores.}
\label{prot_polite}
\end{table}
\subsection*{A3. Illustration that Assumption~\ref{as3} is not always satisfied exactly}
\begin{figure}[h!]
    \includegraphics[width=0.8\textwidth]{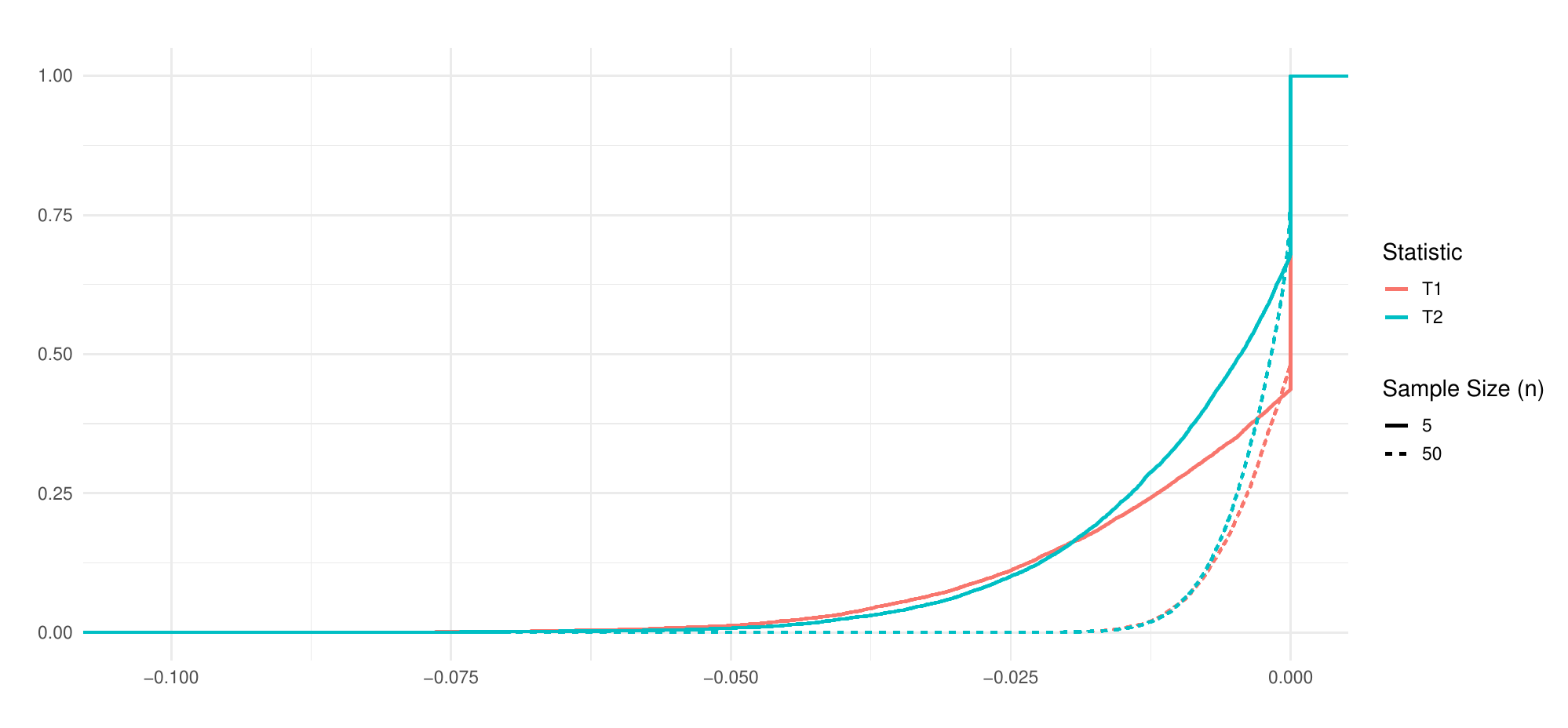}
    \caption{Assumption~\ref{as3}, i.e., $ T(X^{(n)},Y^{(n)}) \succsim_{FSD} \tilde{T}(X^{(n)},\tilde{Y}^{(n)})$ is violated.}\label{figure_assumption_3_ssd}
    \end{figure}

For example for classical second order stochastic dominance, Assumption~\ref{as3} is generally only asymptotically valid. For example, take $Y \sim \text{Exp}(20) +20$ and $X\sim \text{Exp}(20) + 20 + \varepsilon$ (independently from $Y$), where $\varepsilon\sim \text{Exp}(1)-1$ is noise (independent form $(X,Y)$) with expectation $0$. Therefore $X$ is a mean preserving spread of $Y$ and thus $X\precsim_{SSD}Y$.
Now, take $X_1,\ldots,X_n \stackrel{i.i.d}{\sim} X$ and $Y_1,\ldots,Y_n \stackrel{i.i.d}{\sim} Y$ independently from $X_1,\ldots,X_n$ and compute the distribution of the test statistics $T_1:=T(X^{(n)},Y^{(n)})$ and $T_2:=\tilde{T}(X^{(n)},\tilde{Y}^{(n)})$, cf., Assumption~\ref{as3}. Figure~\ref{figure_assumption_3_ssd} shows the distribution of the corresponding test statistic $T_1$ and the permutet test statistic $T_2$ for $n=5$ and $n=50$, where $\mathcal{F}:=\{u : \mathbb{R} \longrightarrow [0,1], u \mbox{ concave }\}$. We realized the test statistic via numerically integrated weighted empirical distribution functions, see the R code. Concretely, our test statistic $T_1$ is very similar (but not exactly identical) to $ -\sqrt{\frac{NM}{N+M}} \hat{S}_2$ in \citet[p. 75]{Barrett}. (But also the statistic of \citet{Barrett} does not excatly satisfy Assumption~\ref{as3}, cf., the R code.)
\newpage
\lstset{
  language=R,
  basicstyle=\ttfamily\small,
  columns=fullflexible,
  breaklines=true,
  linewidth=0.9\textwidth
}
\begin{lstlisting}[language=R,breaklines=true, breakatwhitespace=true]

n_rep <-10000
T_1_5 <- T_2_5 <- T_1_50 <- T_2_50 <- rep(0,n_rep)

T <- function(x,y,grid=seq(0,40,length.out=50000)){
     # test statistic that tests for second order stochastic dominance,
	 # concretely H_0: F_x <=_{SSD} F_y
	 # negative values of T point towards rejection of H_0
	 # T is similar, but not identical to the test statistic \hat{S}_2 in Bartett & Donald 2003, # p. 75, we have T(x,y) \approx (-1)* sqrt(N+M/(M*N))* \hat{S}_2 (where \hat{F} corresponds to the cdf of x and \hat{G} corresponds to the cdf of y in Bartlett, respectively 
     #	 
     grid_min <- grid[1]
	 grid_max <- grid[length(grid)]
	 if(any(x < grid_min | y < grid_min | x > grid_max | y > grid_max)){
	   print("warning: grid too small")
	   
	   return(NULL)
	 }
	 F_x <- ecdf(x)
	 F_y <- ecdf(y)
	 m <- length(x)
	 n <- length(y)
	 statistic <- (max(grid)-min(grid))/length(grid)*
	 min(1/(grid[-1])*(cumsum(F_x(grid[-1]))-cumsum(F_y(grid[-1]))))
	  # The statistic in Bartett &Donald 2003 would be:
	  # statistic  <- (-1)*-sqrt(n*m/(n+m)(max(grid)-min(grid))/length(grid)
	  # *min((cumsum(F_x(grid[-1]))-cumsum(F_y(grid[-1]))))
	  # The main difference is that the weighting 1/(grid[-1]) iin the min is missing
	  
return(statistic)}# *

   
for(k in (1:n_rep)){
   
   # X >_SD2 Y
   # testing H0: X <=_{SSD} Y 
   # reject if T < c_crit
   
   # n=5:
   epsilon <- rexp(5,rate=1)- 1 # noise with expectation 0
   y <- rexp(5,rate=20) + 20
   x <- rexp(5,rate=20) + 20 + epsilon # x is a mean-preserving spread of y,
										# therefore X 	<_SSD Y 
   z <- sample(c(x,y))  # pooled sample
   T_1_5[k] <- T(x,y)
   T_2_5[k] <- T(z[(1:5)],z[-(1:5)]) 	

   # n=50:
   epsilon <- rexp(50,rate=1)- 1 # noise with expectation 0
   y <- rexp(50,rate=20)  + 20
   x <- rexp(50,rate=20)  + 20 + epsilon # x is a mean-preserving spread of y,
										# therefore X 	<_SSD Y 
   z <- sample(c(x,y))  # pooled sample
   T_1_50[k] <- T(x,y)
   T_2_50[k] <- T(z[(1:50)],z[-(1:50)]) 	   
   
}
   
# produce graphics
# Load required library
library(ggplot2)

# Combine all data into one data frame for ggplot
df <- data.frame(
  value = c(T_1_5, T_2_5, T_1_50, T_2_50),
  statistic = factor(rep(c("T1", "T2", "T1", "T2"), 
                         times = c(length(T_1_5), length(T_2_5), length(T_1_50), length(T_2_50)))),
  n = factor(rep(c(5, 5, 50, 50), 
                 times = c(length(T_1_5), length(T_2_5), length(T_1_50), length(T_2_50))))
)

# Plot cdfs of all test statistics
ggplot(df, aes(x = value, color = statistic, linetype = n)) +
  stat_ecdf(size = 1) +
  labs(
    title = "",
    x = "",
    y = "",
    color = "Statistic",
    linetype = "Sample Size (n)"
  ) +
  #xlim(c(-1.75,0))+
  theme_minimal(base_size = 14) +
  theme(
    legend.position = "right",
    plot.title = element_text(hjust = 0.5)
  )
 \end{lstlisting}
\bibliographystyle{apalike}
\bibliography{references.bib}
\end{document}